\documentclass[amsmath, amssymb, preprintnumbers, showpacs, showkeys, aps, prb,superscriptaddress,twocolumn]{revtex4-2}
\usepackage{amssymb}
\usepackage[utf8]{inputenc}
\usepackage{graphicx}
\usepackage{amsmath}
\usepackage{csquotes}
\usepackage{color, xcolor}
\usepackage{placeins}
\usepackage{braket}
\usepackage{comment}
\usepackage[colorlinks=true, linkcolor={red!80!black},citecolor={blue!70!black},urlcolor={blue!80!black},
pdfstartview=FitV,
pdftitle={},
pdfauthor={},
pdfsubject={},
pdfkeywords={},
pdfpagemode=None,
bookmarksopen=true
]{hyperref}

\newcommand{\beq}{\begin{equation}}
	\newcommand{\eneq}{\end{equation}}

\newcommand{\ii}{\text{i}}

\usepackage{amsmath, amssymb, ascmac, mathtools, bm, braket, bbold}

\usepackage{ulem} 
\begin{document}

\title{Hinge modes of three-dimensional Euler insulators}
	
\author{Yutaro Tanaka}
%\email{yutaro.tanaka.ay@riken.jp}
\affiliation{RIKEN Center for Emergent Matter Science, Wako, Saitama, 351-0198, Japan}

\author{Shingo Kobayashi}
%\email{shingo.kobayashi@riken.jp}
\affiliation{RIKEN Center for Emergent Matter Science, Wako, Saitama, 351-0198, Japan}

%\author{Name}
%\affiliation{affiliation}

%\date{\today}
	
\begin{abstract} 
In two-dimensional systems with space-time inversion symmetry, such as $C_{2z}\mathcal{T}$, the reality condition on wave functions gives rise to real band topology characterized by the Euler class, a $\mathbb{Z}$-valued topological invariant for a pair of real bands in the Brillouin zone. 
In this paper, we study three-dimensional $C_{2z}\mathcal{T}$-symmetric insulators characterized by $\bar{e}_2$, defined as the difference in the Euler classes between two $C_{2z}\mathcal{T}$-invariant planes in the three-dimensional Brillouin zone. 
By deriving effective surface Hamiltonians from generic low-energy continuum Hamiltonians characterized by the topological invariant $\bar{e}_2$, we reveal that multiple gapless boundary states exist at the domain walls of the surface mass, which give rise to the multiple chiral hinge modes.
We also show that three-dimensional insulators characterized by $\bar{e}_2=N$ support $N$ chiral hinge modes. Notably, due to the constraint of two occupied bands in our system, these phases are distinct from stacked Chern insulators composed of $N$ layers.
Furthermore, we construct tight-binding models for $\bar{e}_2=2$ and $3$ and numerically demonstrate the emergence of two and three chiral hinge modes, respectively. These results are consistent with those obtained from the surface theory.
\end{abstract}
	
\maketitle
\section{Introduction}
The discovery of topological insulators established a new paradigm in condensed matter physics \cite{RevModPhys.82.3045, RevModPhys.83.1057}. 
A defining hallmark of topological insulators is the bulk-boundary correspondence, which stipulates that $d$-dimensional systems possess $(d-1)$-dimensional boundary states characterized by bulk topological invariants. This concept extends naturally to higher-order topological insulators, in which $d$-dimensional $n$th-order topological phases support $(d-n)$-dimensional boundary states~\cite{PhysRevLett.108.126807, PhysRevLett.110.046404, benalcazar2017quantized, PhysRevB.96.245115, PhysRevLett.119.246401, PhysRevLett.119.246402, schindler2018higher, fang2017rotation}. 
Driven by this generalized bulk-boundary correspondence, higher-order topological phases have attracted significant research interest in recent years~\cite{PhysRevB.97.205135, PhysRevB.97.241405, schindler2018higherbismuth, PhysRevB.98.205147, serra2018observationnature555, peterson2018quantizedNature7695, imhof2018topolectricalnatphys, PhysRevLett.123.016806, wang2018higher, sheng2019two,  PhysRevB.98.035147, okugawa2019second, ghosh2019engineering, agarwala2019higher, chen2019higher, hirayama2020higher, chen2020universal, physrevb.101.115120, physrevresearch.2.013300,  PhysRevResearch.2.043274, PhysRevResearch.3.L032029, PhysRevResearch.3.033177, PhysRevLett.127.176601,   PhysRevB.103.115118,  PhysRevB.104.245427,  PhysRevB.105.115119, physrevb.107.245148, Yamazaki2025, Tanaka2025PRB, tanaka2025}.

A systematic method for classifying these topological phases is the $K$-theory approach~\cite{Schnyder2008PRB, Kitaev2009, Ryu2010NJP}, wherein the nontrivial topology is stable under the addition of topologically trivial bands.
However, certain topological phases fall outside this classification framework.
In both spinless and spinful two-dimensional systems with $C_{2z}\mathcal{T}$ symmetry, where $C_{2z}$ and $\mathcal{T}$ denote a two-fold rotation about the $z$-axis and the time-reversal operator, respectively, one can always choose a real gauge in which both the Hamiltonian and the wave functions are real~\cite{physrevb.91.161105, Ahn2018PRL, Ahn2019CPB}, giving rise to real band topology~\cite{Morimoto2014PRB, Fang2015PRB,  Zhao2017PRL, Bzduifmmode2017PRB, Bouhon2019PRB, BouhonPRB2020, Brouwer2023PRB}.
In systems with real bands, a $\mathbb{Z}$-valued topological invariant $e_2$, known as Euler class \cite{Ahn2019PRX,  hatcher_VBKT}, characterizes exotic topological phenomena associated with a pair of real bands isolated from the rest of the bands, such as the violation of the fermion doubling theorem and non-Abelian braiding of band degeneracy points  \cite{Ahn2019PRX, Wu2019Science, Bouhon2020}.
Since the Euler class is a multi-gap invariant defined for a specific number of bands isolated from others, the Euler band topology falls outside the established classification theory of topological phases that are stable under the addition of trivial bands. 

Despite extensive previous studies on the Euler band topology~\cite{Unal2020PRL,  Ezawa2021PRB, Guan2022PRR, Chen2022PRB, Slager2024Floquet, Ghadimi2024PRL, Davoyan2024PRB, Jankowski2024PRB, JankowskiPRB2024, Sato2024, Lee2025PRB, Kobayashi2026PRB, Mondal2026PRB, Karle2026PRA}, its role in three-dimensional (3D) insulators remains elusive.
In this paper, we focus on $C_{2z}\mathcal{T}$-symmetric 3D band insulators exhibiting a difference in the Euler class between two $C_{2z}\mathcal{T}$-invariant planes in the 3D Brillouin zone, which we refer to as 3D Euler insulators.
Previous works have elucidated that such 3D Euler insulators support characteristic surface states on the $C_{2z}\mathcal{T}$-invariant surface, namely, the (001) surface~\cite{Kobayashi2021PRB, Sato2025PRB}.
Boundary and hinge modes associated with band topology defined on planes of the 3D Brillouin zone, including the Euler class, have also been discussed within the framework of subdimensional topology~\cite{Bouhon2021PRB, Lange2021PRB}. However, a systematic correspondence between the value of the Euler class and the number of chiral hinge modes has remained unexplored.

In this paper, based on generic low-energy continuum theories and tight-binding models, we show that 3D Euler insulators characterized by the topological invariant $\bar{e}_2=e_2(0)-e_2(\pi)$, defined as the difference in the Euler class between the $k_z=0$ and $k_z=\pi$ planes, support multiple one-dimensional hinge modes along the $z$ direction. 

\begin{figure}
\includegraphics[width=1.\columnwidth]{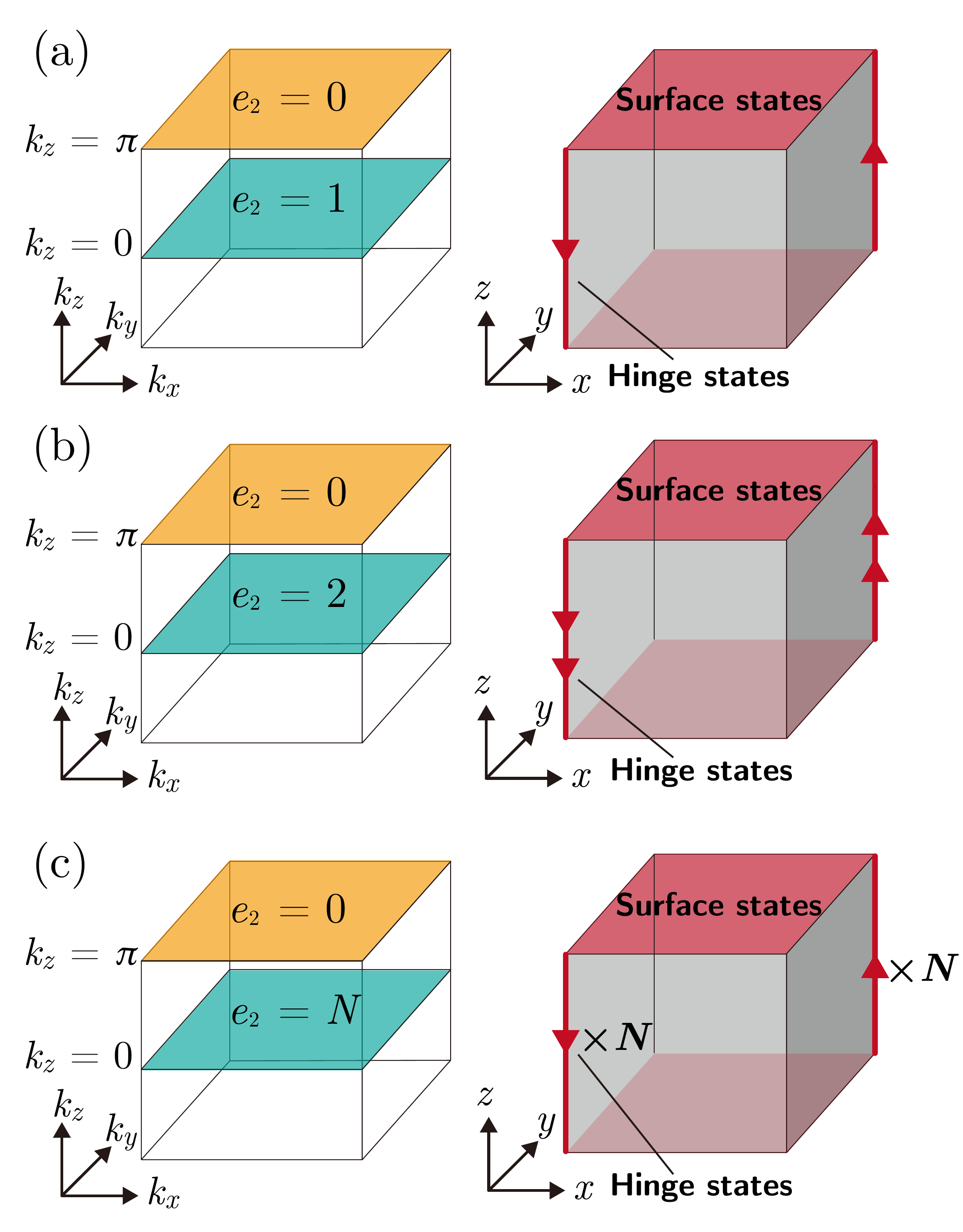}
	\caption{Three-dimensional Euler insulators with multiple hinge modes. The left panels show schematics of the Euler class $e_2(k_z)$ in the $k_z=0$ and $k_z=\pi$ planes within the Brillouin zone for (a) $\bar{e}_2:=e_2(0)-e_2(\pi)=1$, (b) $\bar{e}_2=2$, and (c) $\bar{e}_2=N$. The right panels depict the real-space configurations of the corresponding Euler insulators with multiple hinge modes.
 }
	\label{fig:concept}
\end{figure}

We also elucidate that the number of hinge modes corresponds to the value of the topological invariant $\bar{e}_2$~[Fig.~\ref{fig:concept}(a-c)]. 
By deriving effective surface Hamiltonians from the generic continuum Hamiltonians exhibiting a nonzero Euler class, we reveal that multiple gapless solutions exist at the domain walls of the surface mass, which bind one, two, and three chiral hinge modes for $\bar{e}_2=1$, $2$, and $3$, respectively. Furthermore, we generalize our continuum theory to arbitrary $\bar{e}_2=N$ and show that 3D Euler insulators support $N$ chiral hinge modes for positive integer $N$.

We numerically demonstrate the emergence of these chiral hinge modes using two tight-binding models corresponding to the continuum Hamiltonians with the topological invariants $\bar{e}_2=2$ and $\bar{e}_2=3$, confirming agreement with the predictions of the continuum theory.

This paper is organized as follows. In Sec.~\ref{sec: Euler_class_review}, we review the reality condition on wave functions in the presence of $C_{2z}\mathcal{T}$ symmetry and the Euler class, which characterizes the band topology associated with a pair of real bands isolated from the rest of the bands. 
In Sec.~\ref{sec: e2=1}, we review that 3D Euler insulators characterized by $\bar{e}_2=1$ host a single chiral hinge mode. 
In Sec.~\ref{sec: e2=2}, we show that 3D Euler insulators characterized by $\bar{e}_2=2$ host double chiral hinge modes using both a generic low-energy continuum theory and numerical calculations of the corresponding tight-binding model. 
In Sec.~\ref{sec: e2=3}, we demonstrate that Euler insulators characterized by $\bar{e}_2=3$ support triple chiral hinge modes, employing a similar approach based on a continuum theory and tight-binding calculations.
In Sec.~\ref{sec: e2=N}, by generalizing the low-energy continuum theory for $\bar{e}_2=1$, $2$, and $3$ in Secs.~\ref{sec: e2=1}, \ref{sec: e2=2}, and \ref{sec: e2=3}, we show that a 3D Euler insulator characterized by $\bar{e}_2=N$ supports $N$ hinge modes, where $N$ is a positive integer. 
Finally, conclusions and discussion are given in Sec.~\ref{sec: conclusions_discussion}.

\section{Euler band topology in 3D insulators}\label{sec: Euler_class_review}
In this section, we review the Euler band topology in 3D insulators. 
In this work, we focus on $C_{2z}\mathcal{T}$-symmetric systems satisfying
\begin{align}\label{eq: C2T_symm}
        C_{2z}\mathcal{T}\mathcal{H}(\boldsymbol{k}_{\parallel}, k_{z}^*)(C_{2z}\mathcal{T})^{-1}=\mathcal{H}(\boldsymbol{k}_{\parallel},k_z^*),
\end{align}
where $C_{2z}$ denotes two-fold rotation about the $z$ axis, $\mathcal{T}$ denotes the time-reversal operator, $\boldsymbol{k}_{\parallel}:=(k_x,k_y)$, and $k_{z}^{*}\in \{0,\pi \}$.
Throughout this work, we consider systems possessing only $C_{2z}\mathcal{T}$ and translational symmetries.
Since $C_{2z}\mathcal{T}$ is an anti-unitary operator satisfying $(C_{2z}\mathcal{T})^2=1$ for both spinless ($\mathcal{T}^2=1$) and spinful ($\mathcal{T}^2=-1$) systems, a real gauge can be imposed on the wave function $\ket{u_n(\boldsymbol{k}_{\parallel},k_z^{*})}$:
\begin{align}
    C_{2z}\mathcal{T}\ket{u_n(\boldsymbol{k}_{\parallel},k_z^{*})} = \ket{u_n(\boldsymbol{k}_{\parallel},k_z^{*})}.
\end{align}
Here, we choose the symmetry representation $C_{2z}\mathcal{T}=K$ without loss of generality, where $K$ denotes complex conjugation. 
We employ this real gauge throughout this work.

For a system with two occupied bands, we can characterize the band topology in the $C_{2z}\mathcal{T}$-invariant plane (the $k_x$-$k_y$ plane at $k_z=k_z^{*}$) by the Euler class $e_2$.
The Euler class $e_2$ is given by the flux integral 
\begin{align} \label{eq: eular}
    e_2(k_z^{*})=\frac{1}{2\pi}\int_{\textrm{2D}}d\boldsymbol{S} \cdot \tilde{\boldsymbol{F}}_{12}(\boldsymbol{k}_{\parallel},k_z^{*}), 
\end{align}
where the integral is taken over the $k_z=k_z^{*}$ plane in the 3D Brillouin zone, $\tilde{\boldsymbol{F}}_{mn}:=\boldsymbol{\nabla}_{\boldsymbol{k}}\times \tilde{\boldsymbol{A}}_{mn}(\boldsymbol{k})$ is the real Berry curvature, and $\tilde{\boldsymbol{A}}_{mn}(\boldsymbol{k}):=\braket{u_m|\boldsymbol{\nabla}_{\boldsymbol{k}} u_n}$ ($m,n \in \{1,2 \}$) is the real Berry connection defined by the real occupied states $\ket{u_n}$. The Euler class is well defined only for orientable wave functions, and the systems considered below satisfy this orientability condition. More precisely,  the first Stiefel–Whitney class $w_1$ characterizes the orientability of real Bloch wave functions. In the presence of $C_{2z}\mathcal{T}$ symmetry, $w_1$ is equivalent to the quantized Berry (Zak) phase. Since the Euler class can be defined only for an orientable real vector bundle, one must require $w_1=0$. Thus, throughout the following discussion, we consider only models with $w_1=0$, for which the real Bloch wave functions are orientable.

Here, we introduce the difference in the Euler class $e_2$ between the $k_z=0$ and $k_z=\pi$ planes,
\begin{align}
    \bar{e}_{2}:=e_2(0)-e_2(\pi),
\end{align}
as a topological invariant characterizing a pair of isolated bands (or when the number of occupied bands is two) in 3D $C_{2z}\mathcal{T}$-symmetric insulators. 
This definition is motivated by various previous works~\cite{PhysRevLett.106.106802, Kobayashi2021PRB,  Ahn2019PRB, Sato2025PRB, Kobayashi2026PRB}, in which the real band topology of $C_{2}\mathcal{T}$-symmetric systems is characterized by the difference in 2D topological invariants between the $k_z=0$ and $k_z=\pi$ planes.

We note that, in $P\mathcal{T}$ symmetric systems, where $P$ denotes spatial inversion, the difference in the Euler class between the $k_z=0$ and $k_z = \pi$ planes is also related to the linking structure of bulk nodal lines~\cite{Ahn2018PRL, Ahn2019CPB,bouhon2022multi,Sato2024}. In contrast, the present work focuses on fully gapped 3D insulators protected only by $C_{2z}\mathcal{T}$ and translational symmetries, where no such nodal structure is present. We establish a bulk-boundary correspondence directly between $\bar{e}_2 = N$ and the number $N$ of chiral hinge modes.

In particular, when $\bar{e}_2=1$, it has been shown that the system supports a single chiral hinge mode, since $\bar{e}_2 \mod 2$ is equivalent to the Chern-Simons invariant~\cite{Ahn2019PRB}. In addition, since the Chern-Simons invariant in 3D $C_{2z}\mathcal{T}$-symmetric insulators is a stable $\mathbb{Z}_2$ topological invariant defined for an arbitrary number of occupied bands, the single chiral hinge mode remains stable upon adding trivial bands~\cite{Ahn2019PRB}.
More precisely, this stable $\mathbb{Z}_2$ invariant is the second Stiefel--Whitney class $w_2 = \bar{e}_2 \bmod 2$, i.e. the parity of the Euler class. Accordingly, for odd $\bar{e}_2$ the single chiral hinge mode is protected by the nontrivial $w_2$, whereas for even $\bar{e}_2$ one has $w_2 = 0$, so that the chiral hinge modes discussed below cannot be attributed to $w_2$ and instead originate intrinsically from the
$\mathbb{Z}$-valued Euler class.

In what follows, we extend this relation to $\bar{e}_2\geq 2$. 
We assume that the number of occupied bands is fixed to two. Since $e_2$ is a fragile topological invariant, the hinge states discussed below are valid only for two occupied bands and become unstable under the addition of bands, for example, by attaching 2D Chern insulators to the three-dimensional surface. 

In calculating $\bar{e}_2$, there are two subtleties. First, the sign of $\bar{e}_2$ is not gauge invariant, since the signs of $e_2(k_z^\ast)$ flip under an $O(2)$ transformation with determinant $-1$. Second, the evaluation of $\bar{e}_2$ requires fixing the relative sign between $e_2(0)$ and $e_2(\pi)$~\cite{Sato2025PRB}.
To circumvent both subtleties, we place the nontrivial Euler class on a single plane by assuming (i) $e_2(0)\neq 0$ and $e_2(\pi)=0$. The invariant then reduces to $\bar{e}_2 = e_2(0)$, whose magnitude $|\bar{e}_2| = |e_2(0)|$ is fixed by one plane alone and is thus free of the relative-sign ambiguity.

\section{Euler class $\bar{e}_{2}=1$}\label{sec: e2=1}

In this section, we revisit the chiral hinge mode appearing in a 3D Euler insulator characterized by $\bar{e}_2=1$ based on a low-energy continuum Hamiltonian.
We show that our approach reproduces the results of previous studies and clarifies the physical origin of the chiral hinge mode.

Let us consider a generic low-energy continuum Hamiltonian
\begin{align}\label{eq: continuum_model_e2=1}
    \mathcal{H}= &v_x k_x \Gamma_1 + v_y k_y\Gamma_2 + v_z k_z \Gamma_3 + \lambda \Gamma_4,
\end{align}
where $v_x$, $v_y$, $v_z$, and $\lambda$ are positive real parameters, and  $\Gamma_i$ ($i=1,\cdots, 4$) are the $4\times 4$ gamma matrices satisfying $\{\Gamma_i,\Gamma_j\}=2\delta_{ij}$.
The symmetry representation is given by $C_{2z}\mathcal{T}=K$, where $K$ denotes complex conjugation.
Since the Hamiltonian $\mathcal{H}$ respects $C_{2z}\mathcal{T}$ symmetry~[Eq.~\eqref{eq: C2T_symm}], the gamma matrices $\Gamma_i$ satisfy 
\begin{gather}
    [C_{2z}\mathcal{T}, \Gamma_1]= 
    [C_{2z}\mathcal{T}, \Gamma_2] = [C_{2z}\mathcal{T}, \Gamma_4]=0, \nonumber \\
    \{ C_{2z}\mathcal{T}, \Gamma_3\}=0.
\end{gather}
The Euler class $|\bar{e}_2|=1$ of this model can be verified by evaluating Eq.~(\ref{eq: eular}) using the occupied states in a real gauge. Here, however, we present an alternative explanation. In evaluating $\bar{e}_2$ in the continuum Hamiltonian, we compactify momentum space by adding a point at infinity, which yields $S^3$, where $k_z=\pi$ corresponds to $k_z=\infty$. Under this compactification, we obtain $e_2(\infty)=0$. Hence, the topology of the system is determined by the Euler class defined on the $S^2$ in the $k_x$-$k_y$ plane at $k_z=0$. On this plane, the third term vanishes. In addition, the fourth term plays the role of a mass term that induces band inversion. Therefore, the value of $e_2$ is determined from the first and second terms, which break both time-reversal and $C_{2z}$ symmetries, where $\mathcal{T}=K$ and $C_{2z}=1$. $|e_2(0)|$ is evaluated by the band inversion and the topology of band touching points (i.e., nodal points) in the occupied bands as follows. 

In real two-band systems with a nontrivial Euler class, nodal points appear in the two bands, and the Euler class is related to a one-dimensional winding number around the nodal points as~\cite{Ahn2019PRX}
\begin{equation}\label{eq: Euler_winding}
e_2 = -\frac{N_{\rm t}}{2},
\end{equation}
where $N_{\rm t}$ denotes the total winding numbers on $S^2$. In the present case, the first and second terms vanish at $k_x=k_y=0$ and at infinity due to the compactification, resulting in two nodal points. Since the dispersion around the nodal points is linear, each winding number is one. In the trivial phase, the two nodal points carry opposite signs of the winding number. Thus, $e_2=0$. On the other hand, when band inversion occurs at $k_x=k_y=0$, the sign of the winding number around $k_x=k_y=0$ flips through the braiding of band degeneracy points between occupied and unoccupied bands~\cite{Ahn2019PRX, Sato2025PRB}, which results in $|e_2|=1$. 

\begin{figure}
\includegraphics[width=1.\columnwidth]{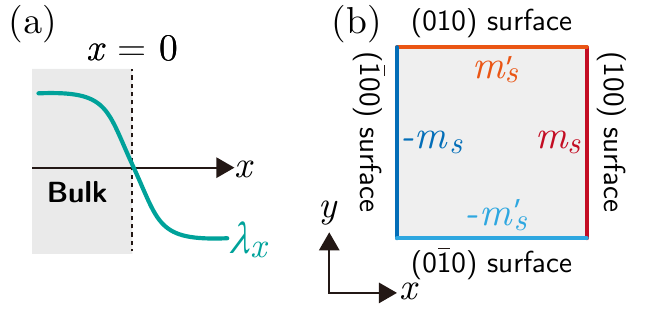}
	\caption{(a)~The spatial profile of $\lambda_x$ around $x=0$ for the ($100$) surface. (b)~The surface mass terms on the ($100$), ($\bar{1}00$), ($010$), and ($0\bar{1}0$) surfaces.
 }
	\label{fig:spatial_profile}
\end{figure}

We derive an effective Hamiltonian for the (100) surface from the bulk Hamiltonian given in Eq.~\eqref{eq: continuum_model_e2=1}.
Following the standard procedure~\cite{Jackiw1976, Zhang2012PRB, physrevb.97.205136, physrevx.8.031070} for deriving surface Hamiltonians, we define $x$ as the coordinate normal to the surface and introduce an $x$-dependence to the coefficient $\lambda$ of the mass term $\lambda \Gamma_4$ via the replacement $\lambda \rightarrow \lambda_x$.
The system is modeled such that the bulk crystal occupies the region $x<0$, the vacuum occupies $x>0$, and the surface is located at $x=0$.
The spatial profile of $\lambda_x$ is defined to vanish at the surface ($\lambda_{x=0}=0$) and to vary sharply towards $\lambda_x=1$ inside the bulk for $x<0$, and towards $\lambda_x=-1$ outside the crystal $x>0$~[Fig.~\ref{fig:spatial_profile}(a)]. 
Consequently, the bulk crystal is characterized by $\lambda_x=1$, while the vacuum is represented by $\lambda_x=-1$.
Replacing $k_x$ with $-\ii \partial_x$, the Hamiltonian in Eq.~\eqref{eq: continuum_model_e2=1} becomes 
\begin{align}
    \mathcal{H}= -\ii v_x \partial_x \Gamma_1 + v_yk_y\Gamma_2 +v_z k_z \Gamma_3 + \lambda_x \Gamma_4. \label{eq: continuum_model_e2=1_2}
\end{align}

The solutions $\psi$ for this Hamiltonian $\mathcal{H}$ localized at $x=0$ are given by
\begin{align}
    \psi = \exp \left( \frac{1}{v_x} \int^{x}_0 dx' \lambda_{x'} \right) P \psi_{k_y,k_z}, \label{eq: solution_e2=1}
\end{align}
where the projection operator $P$ is defined as 
\begin{align}
     P:= \frac{1}{2}\left( 1 - \ii \Gamma_1 \Gamma_4 \right).
\end{align}
Note that the projection operators $P$ satisfy
\begin{gather}
     P^2=P, \quad   [P,\Gamma_2]=[P,\Gamma_3]=0. 
     \label{eq: commutation_e2=1}
\end{gather}
By applying the Hamiltonian in Eq.~\eqref{eq: continuum_model_e2=1_2} to the eigenstate $\psi$ in Eq.~\eqref{eq: solution_e2=1}, we obtain 
\begin{align}
    \mathcal{H}\psi = \exp \left( \frac{1}{v_x} \int^{x}_0 dx' \lambda_{x'} \right) \mathcal{H}_{s} \psi_{k_y,k_z}.
\end{align}
Here, $\mathcal{H}_{s}$ is the effective surface Hamiltonian defined as
\begin{align}
    \mathcal{H}_{s}:=P(v_y k_y \Gamma_2 + v_z k_{z}\Gamma_3)P, \label{eq: surface_bar100_e2=1}
\end{align}
which describes the surface Dirac cone on the $(100)$ surface.

Furthermore, by substituting $\partial_x \rightarrow - \partial_x$ in Eq.~\eqref{eq: continuum_model_e2=1_2}, we obtain the $(\bar{1}00)$ surface Hamiltonian. Following a procedure similar to that used for the $(100)$ surface, we find that this Hamiltonian is identical to Eq.~\eqref{eq: surface_bar100_e2=1}, with the projection operator given by ${P} = \left( 1 + \ii \Gamma_1 \Gamma_4 \right)/2$. Thus, both the $(100)$ and $(\bar{1}00)$ surface Hamiltonians demonstrate the emergence of gapless boundary states on these surfaces.

In addition, we also obtain the $(010)$ and $(0\bar{1}0)$ surface Hamiltonians in a manner analogous to the $(100)$ and $(\bar{1}00)$ surfaces by replacing $k_y$ as $-\ii \partial_y$ for the $(010)$ surface and $\ii \partial_y$ for the $(0\bar{1}0)$ surface in the Hamiltonian presented in Eq.~\eqref{eq: continuum_model_e2=1}.
Through similar procedures to the $(100)$ and $(\bar{1}00)$ surfaces, we obtain the surface Hamiltonian
\begin{align}
    \mathcal{H}_{s}:=P'(v_x k_x \Gamma_1 + v_z k_{z}\Gamma_3)P', \label{eq: surface_010_e2=1}
\end{align}
where the projection operator $P'$ is given by $P':= \left( 1 - \ii \Gamma_2 \Gamma_4 \right)/2$ for the ($010$) surface and by $P':= \left( 1 + \ii \Gamma_2 \Gamma_4 \right)/2$ for the ($0\bar{1}0$) surface.

To see a chiral hinge mode, we introduce a time-reversal-symmetry-breaking mass term that gaps out these gapless boundary states. The only possible mass term capable of opening a gap in these surface states is given by $m_{\boldsymbol{n}}\Gamma_5$, where $\Gamma_5$ is the gamma matrix satisfying $\{ \Gamma_5, C_{2z}\mathcal{T} \}=0$, and $m_{\boldsymbol{n}}$ depends on the surface normal vector $\boldsymbol{n}$.
To preserve the $C_{2z}\mathcal{T}$ symmetry, the spatial profile of the mass term must satisfy $m_{\boldsymbol{n}}=-m_{C_{2z}\boldsymbol{n}}$. Therefore, in a rod geometry extending along the $z$ direction and bounded by the $(100)$, $(\bar{1}00)$, $(010)$, and $(0\bar{1}0)$ surfaces, the $(100)$ and $(\bar{1}00)$ surfaces host mass terms with opposite signs, namely, $m_s$ and $-m_s$~[Fig.~\ref{fig:spatial_profile}(b)].
Similarly, the $(010)$ and $(0\bar{1}0)$ surfaces host mass terms with opposite signs, $m_s'$ and $-m_s'$.
When $m_s$ and $m'_s$ share the same sign, zero-mass lines emerge between the $(100)$ and $(0\bar{1}0)$ surfaces, and between the $(010)$ and $(\bar{1}00)$ surfaces. Conversely, when they have opposite signs, zero-mass lines emerge between the $(100)$ and $(010)$ surfaces, and between the $(\bar{1}00)$ and $(0\bar{1}0)$ surfaces.
These zero-mass lines act as domain walls that bind chiral hinge states~\cite{PhysRevLett.108.126807, PhysRevLett.110.046404, PhysRevLett.119.246401, PhysRevLett.119.246402, physrevb.97.205136, physrevx.8.031070, schindler2018higher}.

\section{3D Euler insulators with $\bar{e}_{2}=2$}\label{sec: e2=2}

\subsection{Continuum theory}\label{sec: e2_2_continuum}
Now, we extend the above argument to a 3D Euler insulator characterized by $\bar{e}_2=2$ and show that double chiral hinge modes appear. We start from a generic low-energy continuum Hamiltonian
\begin{align}\label{eq: continuum_model_e2=2}
        \mathcal{H}=&~v_1(k_x^2-k_y^2)\Gamma_1 +v_2k_xk_y\Gamma_2+v_zk_z\Gamma_3 \nonumber \\
        &+(v_{xz}k_x+v_{yz}k_y)k_z\Gamma_4+\lambda \Gamma_5, 
\end{align}
where $v_1$, $v_2$, $v_z$, $v_{xz}$, $v_{yz}$, and $\lambda$ are positive real values, and $\Gamma_i$ ($i=1,\cdots, 5$) are the $4\times 4$ gamma matrices satisfying $\{\Gamma_i,\Gamma_j\}=2\delta_{ij}$.
The symmetry representation is given by $C_{2z}\mathcal{T}=K$, where $K$ denotes complex conjugation.
Since the Hamiltonian $\mathcal{H}$ respects $C_{2z}\mathcal{T}$ symmetry~[Eq.~\eqref{eq: C2T_symm}], the gamma matrices $\Gamma_i$ satisfy
\begin{gather}
    [C_{2z}\mathcal{T}, \Gamma_1]= 
    [C_{2z}\mathcal{T}, \Gamma_2] = [C_{2z}\mathcal{T}, \Gamma_5]=0, \nonumber \\
    \{ C_{2z}\mathcal{T}, \Gamma_3\}=\{ C_{2z}\mathcal{T}, \Gamma_4\}=0.\label{eq: commute_C2T_Gamm}
\end{gather}
The Euler class $|\bar{e}_2|=2$ can be understood in a manner similar to the $|\bar{e}_2|=1$ case. On the $k_z=0$ plane, the third and fourth terms vanish, and the fifth term serves as a mass term responsible for a band inversion. Consequently, the first and second terms determine the value of $e_2$. Since the coefficients of these terms exhibit quadratic dispersion around $k_x=k_y=0$, the corresponding winding number becomes two~\cite{Kobayashi2021PRB}.
Thus, this Hamiltonian exhibits $|\bar{e}_2|=2$ via the band inversion at $k_x=k_y=0$. 
Here, we add the fourth term that breaks both time-reversal symmetry and $C_{2z}$ symmetry ($\mathcal{T}=K$ and $C_{2z}=1$), which plays an important role as a surface mass term in the following argument. 
We note that introducing this winding structure of both the first and the second terms is equivalent to rendering the Euler class nontrivial. A nonzero Euler class is reflected in a nontrivial winding of the band nodes [see Eq.~\eqref{eq: Euler_winding}]. Therefore, incorporating the winding is not an arbitrary choice but an intrinsic requirement for realizing Euler topology. Once this winding structure is imposed, the continuum Hamiltonian compatible with $C_{2z}\mathcal{T}$ symmetry up to second order in momentum is fixed to the form of Eq.~\eqref{eq: continuum_model_e2=2}.

We derive an effective Hamiltonian for the ($100$) surface from the bulk Hamiltonian in Eq.~\eqref{eq: continuum_model_e2=2} in a manner analogous to that in Sec.~\ref{sec: e2=1}, via the replacement $\lambda \rightarrow \lambda_x$. 
The spatial profile of $\lambda_x$ is defined such that it vanishes at the surface ($\lambda_0=0$) and varies sharply from $\lambda_x=0$ to $\lambda_x=1$ for $x<0$, and to $\lambda_x=-1$ for $x>0$. 
By replacing $k_x$ as $-\ii \partial_x$ and setting $v_2k_y=v_{xz}k_z$ or $v_2k_y=-v_{xz}k_z$, the Hamiltonian in Eq.~\eqref{eq: continuum_model_e2=2} reduces to the following Hamiltonian $\mathcal{H}_{i}$ ($i=1,2$):
\begin{align}
    \mathcal{H}_{i}=&-\ii \sqrt{2}v_{xz} k_z \partial_x \tilde{\Gamma}_i-v_1 \partial_x^2 \Gamma_1 +v_z k_z \Gamma_3 +\lambda_x \Gamma_5, 
    \label{eq:surface_dirac_Hamiltonian2}
\end{align}
where $\mathcal{H}_1$ and $\mathcal{H}_{2}$ correspond to $v_2k_y=v_{xz}k_z$ and  $v_2k_y=-v_{xz}k_z$, respectively, and we neglect terms quadratic in the wavevector. Here, $\tilde{\Gamma}_i$ ($i=1,2$) are defined as $\tilde{\Gamma}_1=(\Gamma_2+ \Gamma_4)/\sqrt{2}$ and $\tilde{\Gamma}_2=(-\Gamma_2+ \Gamma_4)/\sqrt{2}$.

The solutions $\psi_{1}$ and $\psi_{2}$ for the Hamiltonians $\mathcal{H}_1$  and $\mathcal{H}_{2}$, respectively, localized at $x=0$ are given by
\begin{align}
    &\psi_{i}=\exp \left( \frac{1}{\sqrt{2}v_{xz}k_z}\int^{x}_{0}dx'\lambda_{x'}\right) P_{i} \psi_{k_z, i},
    \label{eq: solutions}
\end{align}
with $i=1,2$, where the projection operators $P_{i}$ are defined as 
\begin{gather}
P_{i}:=\frac{1}{2}\left( 1 - {\ii} \tilde{\Gamma}_i \Gamma_5 \right)\quad (i=1,2).
    \label{eq:projection_operator}
\end{gather}
Note that the projection operators $P_{i}$ ($i=1,2$) satisfy 
\begin{gather}
     P_i^2=P_i, \quad   [P_i,\Gamma_1]=[P_i,\Gamma_3]=0. 
     \label{eq: commutation_e2}
\end{gather}
Because $[P_1, P_2]\neq0$, $\psi_1$ and $\psi_2$ represent two distinct boundary state solutions.
Equation (\ref{eq: solutions}) diverges at $k_z=0$, which is an artifact of the simplified model.

Applying the Hamiltonian $\mathcal{H}_{i}$ ($i=1,2$) to the states $\psi_{i}$, we obtain
\begin{align}
    \mathcal{H}_{i} \psi_{i}=\exp \left( \frac{1}{\sqrt{2}v_{xz}k_z}\int^{x}_{0}dx'\lambda_{x'}\right)\mathcal{H}_{s,i} \psi_{k_z,i}. \label{eq: H_psi}
\end{align}
Here, $\mathcal{H}_{s,i}$ is the effective surface Hamiltonian corresponding to the (100) surface defined as
\begin{align}
    \mathcal{H}_{s,i}:=P_i( v_z k_{z}\Gamma_3 + m_s \Gamma_1)P_i, \label{eq: surface_100}
\end{align}
with $m_s:= -{v_1 \partial_x \lambda_x}/({\sqrt{2}v_{xz}k_z})$,
where we have neglected terms of order $\mathcal{O}(\lambda_x^2)$ by assuming $\lambda_x \ll 1$ near the surface.
The corresponding energy eigenvalues of the surface Hamiltonian are given by 
\begin{align}
    E=\pm\sqrt{(v_zk_z)^2+m_s^2}. \label{eq: energyeigenvalues}
\end{align}

Via the replacement $\partial_x \rightarrow-\partial_x$ in Eq.~\eqref{eq:surface_dirac_Hamiltonian2}, we can derive the effective surface Hamiltonian corresponding to the ($\bar{1}00$) surface. 
Through a similar procedure to the $(100)$ surface, we obtain the solutions 
\begin{align}
    &\bar{\psi}_{i}=\exp \left( \frac{1}{\sqrt{2}v_{xz}k_z}\int^{x}_{0}dx'\lambda_{x'}\right) \bar{P}_{i} \bar{\psi}_{k_z, i},
    \label{eq: solutions_bar100}
\end{align}
and the surface Hamiltonian for the ($\bar{1}00$) surface
\begin{align}
    \bar{\mathcal{H}}_{s,i}=\bar{P}_i( v_z k_{z}\Gamma_3 - m_s \Gamma_1)\bar{P}_i\label{eq: surface_bar100}
\end{align}
with $i=1,2$, where $\bar{P}_{i}$ is defined as 
\begin{gather}
\bar{P}_{i}:=\frac{1}{2}\left( 1 + {\ii} \tilde{\Gamma}_i \Gamma_5 \right) \quad (i=1,2).
\end{gather}
Thus, we find that these two surfaces host mass terms with opposite signs: $\pm m_s \Gamma_1$ in Eqs.~\eqref{eq: surface_100} and \eqref{eq: surface_bar100}. 

Similarly, we can derive the $(010)$ and $(0\bar{1}0)$ surface Hamiltonians by following a procedure similar to that used for the $(100)$ and $(\bar{1}00)$ surfaces, replacing $k_y$ with $-\ii \partial_y$ in the Hamiltonian presented in Eq.~\eqref{eq: continuum_model_e2=2}. 
By the same reasoning applied to the $(100)$ and $(\bar{1}00)$ surface Hamiltonians, the $(010)$ and $(0\bar{1}0)$ surface Hamiltonians also host mass terms with opposite signs.

In a rod geometry extending along the $z$ direction and bounded by the $(100)$, $(\bar{1}00)$, $(010)$, and $(0\bar{1}0)$ surfaces, the $(100)$ and $(\bar{1}00)$ surfaces host mass terms with opposite signs, as do the $(010)$ and $(0\bar{1}0)$ surfaces~[Fig.~\ref{fig:spatial_profile}(b)].
As discussed in Sec.~\ref{sec: e2=1}, with such surface mass terms, zero-mass lines emerge between the surfaces and act as domain walls that bind chiral hinge states.
Furthermore, since there are two boundary state solutions, $\psi_{1}$ and $\psi_{2}$ (or $\bar{\psi}_{1}$ and $\bar{\psi}_{2}$), exactly two chiral hinge states emerge at each hinge.

While the two chiral hinge modes become gapless at $k_z=0$, the argument of the exponential function in the solutions given in Eq.~\eqref{eq: solutions} diverges at this point. 
To stabilize the chiral hinge modes and avoid this divergence, we can add a perturbation $\Delta \mathcal{D}$ ($\Delta \in \mathbb{R}$) to shift the gapless points from $k_z=0$ to $k_z = \pm \Delta/v_z$, where $\mathcal{D}$ is a matrix satisfying  
\begin{align}
    [\Gamma_3, \mathcal{D} ]=\{ \Gamma_1, \mathcal{D} \}= [C_{2z}\mathcal{T}, \mathcal{D} ]=0. 
\end{align}
In the presence of this perturbation, the corresponding energy eigenvalues are given by
\begin{align}
    E=\pm\sqrt{(v_zk_z\pm \Delta)^2+m_s^2}, \label{eq: energyeigenvalues_with_Delta}
\end{align}
resulting in hinge modes with gapless points located at $k_z =\pm \Delta /v_z$.

\subsection{Tight-binding model}\label{sec: e2=2_lattice}

\begin{figure}
\includegraphics[width=1.\columnwidth]{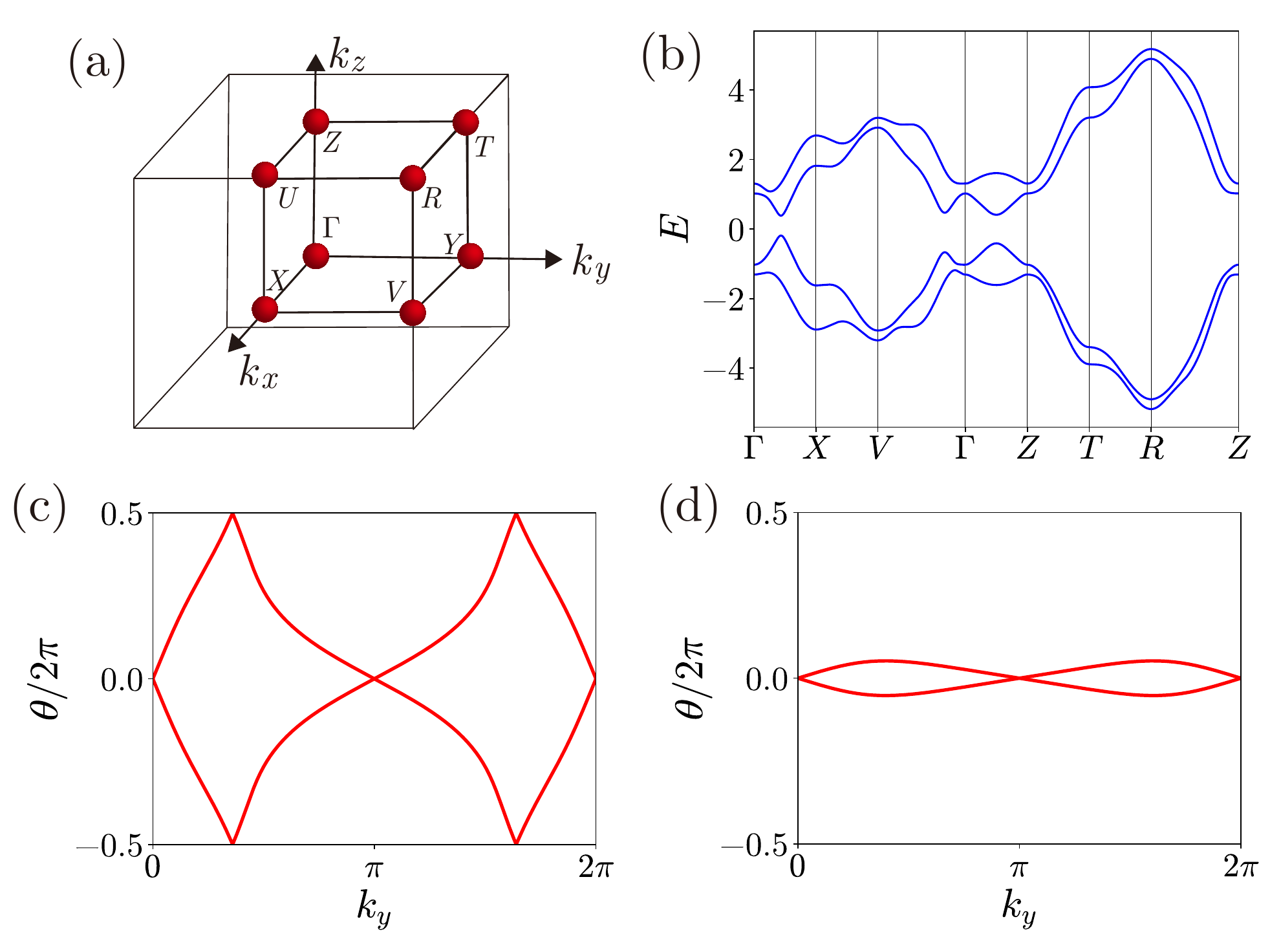}
	\caption{(a)~The Brillouin zone and the high-symmetry points of the model $\mathcal{H}^{(\bar{e}_{2}=2)}_{\boldsymbol{k}}$~[Eq.~\eqref{eq:lattice_model}]. (b) The bulk band structure along the high-symmetry lines. (c,d) The spectra of the $x$-directed Wilson loop matrix at (c) $k_z=0$ and (d) $k_z = \pi$~[$\lambda=1$, $v_1=0.5$, $v_2=2$, $v_{z}=1$, $v_{xz}=v_{yz}=0.5$, $B_1=B_2=0.1$, $\Delta=0.6$]. 
 }
	\label{fig:bulk_EI}
\end{figure}

To numerically verify our theory, we construct a 3D tight-binding model on a simple cubic lattice corresponding to the continuum Hamiltonian in Eq.~\eqref{eq: continuum_model_e2=2}. 
Setting the lattice constant to unity, the lattice sites are denoted by vectors $\boldsymbol{R}=(x,y,z)$ with $x,y,z\in \mathbb{Z}$.
The lattice Hamiltonian is given by
\begin{align}\label{eq:lattice_model}
    \mathcal{H}^{(\bar{e}_{2}=2)}_{\boldsymbol{k}}=&~ 
    2v_{1}\left( \cos k_x -\cos k_y \right) \sigma_x \tau_x +v_{2}\sin k_x \sin k_y \sigma_z \tau_x \nonumber \\ 
    &+\left(v_{xz} \sin k_x + v_{yz}\sin k_y \right) \sin k_z \tau_y \nonumber  \\
     &-\Bigl(3-\lambda-\sum_{i=x,y,z}\cos k_i \Bigr) \tau_z+v_{z} \sin k_z \sigma_y \tau_x  \nonumber \\
     &+B_1 \sigma_x+ B_2 \sigma_z+\Delta \tau_x,
\end{align}
where $\sigma_i$ and $\tau_i$ ($i=x,y,z$) are Pauli matrices. At each lattice site, there are four orbitals labeled as $\ket{\sigma_z}\otimes \ket{\tau_z}$ with $\sigma_z,\tau_z=\pm1$.
This model respects $C_{2z}\mathcal{T}$ symmetry~[Eq.~\eqref{eq: C2T_symm}], where $C_{2z}\mathcal{T}=K$ is the complex conjugation operator.
Throughout this work, we calculate the energy spectrum of the tight-binding models using the PythTB package~\cite{Coh_Vanderbilt_PythTB_2022}.
Figure~\ref{fig:bulk_EI}(a) shows the high-symmetry points in the Brillouin zone. 
Figure~\ref{fig:bulk_EI}(b) indicates the bulk band structures of $ \mathcal{H}^{(\bar{e}_{2}=2)}_{\boldsymbol{k}}$. Setting the Fermi energy to zero, the lower two of the four bands are occupied.
We note that realizing the nonzero Euler class requires hopping terms beyond nearest-neighbor ones, reflecting the higher-order momentum dependence of the winding terms in the continuum Hamiltonian~[Eq.~\eqref{eq: continuum_model_e2=2}]. In particular, the term $v_2 \sin k_x \sin k_y\,\sigma_z\tau_x$ in Eq.~\eqref{eq:lattice_model}, which corresponds to the continuum term $v_2 k_x k_y \Gamma_2$ in Eq.~\eqref{eq: continuum_model_e2=2}, represents diagonal (next-nearest-neighbor) hopping.

To evaluate the Euler class $e_2$ in the $k_x$-$k_y$ plane at $k_z=k_z^{*}$, we employ the Wilson loop matrix~\cite{Soluyanov2011PRB, PhysRevB.84.075119, PhysRevB.89.155114}, defined as 
\begin{align}
	\mathcal{W}(k_{y},k_z) 
	:=   \mathcal{P} \exp \Bigl( \ii \int_{0}^{2\pi} dk_{x} {\mathcal{A}}_x (\boldsymbol{k}) \Bigr), \label{eq: wilson_loop}
\end{align}
where $\mathcal{P}$ denotes the path-ordering operator, and $ {\mathcal{A}}_x (\boldsymbol{k}) $ is the non-Abelian Berry connection matrix with elements
$[{\mathcal{A}}_x (\boldsymbol{k}) ]_{nm}:=\ii \braket{u_{n}(\boldsymbol{k})|\partial_{k_x}u_{m}(\boldsymbol{k})}$.
Since $\mathcal{W}(k_{y},k_z^{*})$ is an SO(2) matrix in the $C_{2z}\mathcal{T}$-invariant plane, the two eigenvalues of $\mathcal{W}(k_{y},k_z^{*})$ are given as a complex conjugate pair $e^{\pm \ii \theta(k_y,k_z^{*})}$, where the phase $ \theta(k_y,k_z^{*})$ corresponds to the Wannier center. The winding number of the evolution of $ \theta(k_y,k_z^{*})$ as $k_y$ varies from $0$ to $2\pi$ is equal to the Euler class $e_{2}$~\cite{Ahn2018PRL, Ahn2019PRX, Bouhon2019PRB, Bouhon2020}. 
Thus, we evaluate the Euler class of the model $\mathcal{H}^{(\bar{e}_{2}=2)}_{\boldsymbol{k}}$ by tracing the evolution of $ \theta(k_y,k_z^{*})$ at $k_z=0$ and $k_z=\pi$~[Figs.~\ref{fig:bulk_EI}(c) and \ref{fig:bulk_EI}(d)]. 
The spectra of the Wilson loop matrix yield $|e_{2}(0)|=2$ and $e_2(\pi)=0$, resulting in the topological invariant $|\bar{e}_2|=2$. 

\begin{figure}
\includegraphics[width=1.\columnwidth]{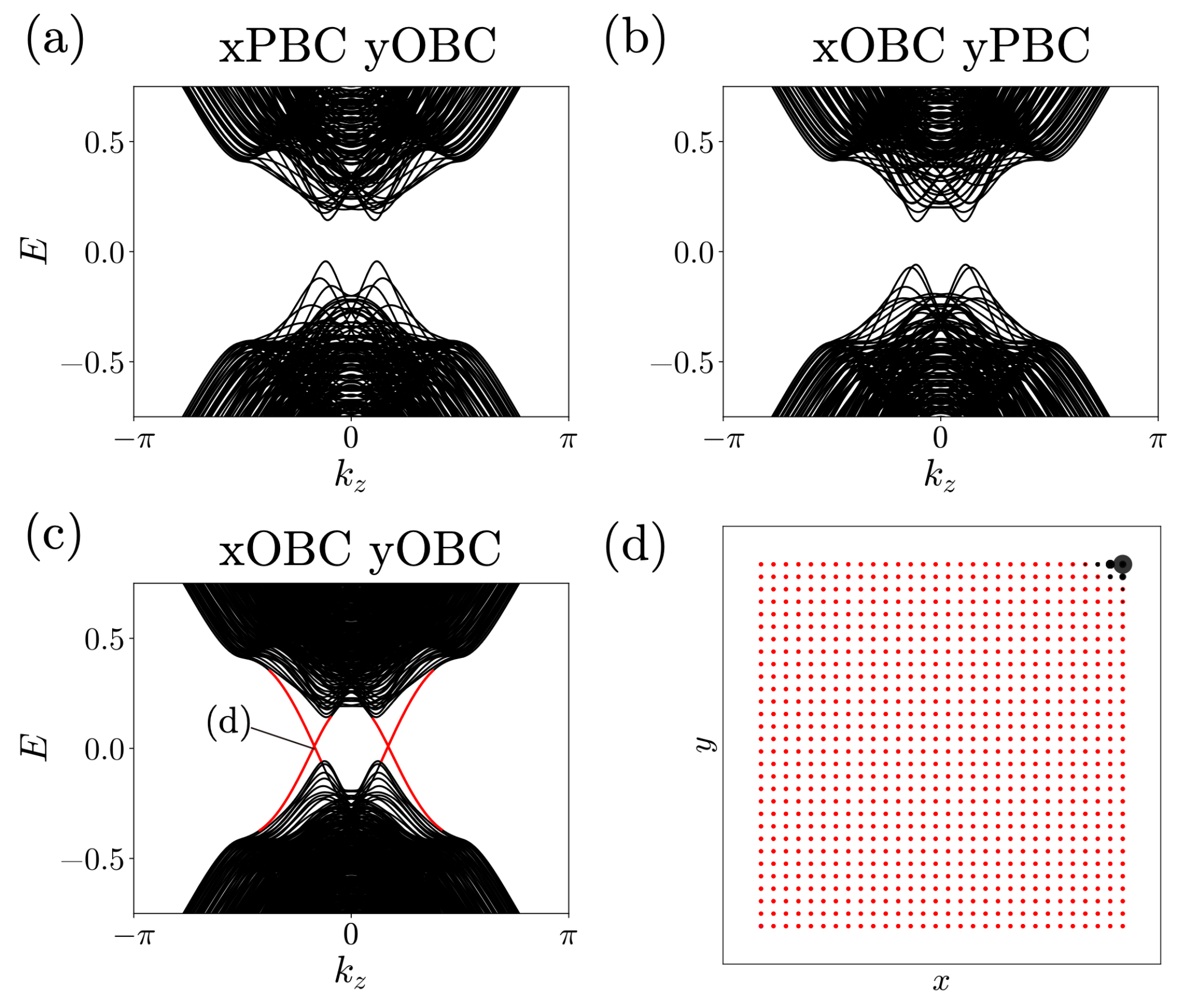}
\caption{(a-c)~Band structures of the model $\mathcal{H}^{(\bar{e}_{2}=2)}_{\boldsymbol{k}}$~[Eq.~\eqref{eq:lattice_model}] along the $k_z$ direction under (a) the periodic boundary conditions (PBCs) both in the $x$ and $z$ directions and the open boundary condition (OBC) in the $y$ direction, (b) the PBCs both in the $y$ and $z$ directions and the OBC in the $x$ direction, and (c) the PBC in the $z$ direction and the OBCs both in the $x$ and $y$ directions. 
(d)~The real-space distribution of the boundary state colored in (c) at ~$k_z=-0.168\pi$ and $E=0.006$. Owing to $C_{2z}\mathcal{T}$ symmetry of the system, equivalent boundary states are localized at the other hinges. Panel (d) shows the distribution at one representative hinge.
The parameter values are the same as those in Fig.~\ref{fig:bulk_EI}. The system sizes in the $x$ and $y$ directions are $L_x=30$ and $L_y=30$, respectively.
}\label{fig:hinge_EI}
\end{figure}

We numerically demonstrate that our model $\mathcal{H}^{(\bar{e}_{2}=2)}_{\boldsymbol{k}}$ supports double chiral hinge modes. As shown in Fig.~\ref{fig:hinge_EI}(a)~(Fig.~\ref{fig:hinge_EI}(b)), this model does not host the gapless boundary states under the periodic boundary conditions (PBCs) along both the $x$ ($y$) and $z$ directions and the open boundary condition (OBC) along the $y$ ($x$) direction. 
Under the PBC along the $z$ direction and the OBCs along the $x$ and $y$ directions, the model supports the double chiral hinge modes at different $k_z$~[Fig.~\ref{fig:hinge_EI}(c)], localized at the hinge~[Fig.~\ref{fig:hinge_EI}(d)].
The emergence of these two hinge modes is consistent with our continuum theory in Sec.~\ref{sec: e2_2_continuum}. 
Since $w_2 = \bar{e}_2 = 0 \bmod 2$ for $\bar{e}_2 = 2$, these two chiral hinge modes cannot be attributed to the second Stiefel--Whitney class. Therefore, they provide numerical evidence that the chiral hinge modes originate from the $\mathbb{Z}$-valued Euler class.
Furthermore, as discussed in Appendix~\ref{appendix: derive_surf_Ham}, we derive an effective surface Hamiltonian from the tight-binding model in Eq.~\eqref{eq:lattice_model}. 
From this analysis, we also confirm that our lattice model supports two hinge modes. 

\section{3D Euler insulators with $\bar{e}_{2}=3$}\label{sec: e2=3}
\subsection{Continuum theory}\label{sec: e2=3_continuum}

In this section, we demonstrate that a 3D Euler insulator characterized by $\bar{e}_2=3$ supports triple chiral hinge modes by considering a generic low-energy continuum Hamiltonian
\begin{align}
    \mathcal{H}=&v_1(k_x^3-3k_x k_y^2)\Gamma_1 + v_2 (3k_x^2 k_y -k_y^3)\Gamma_2 \nonumber \\
    & +v_z k_z \Gamma_3 + ( v_{xz}k_x+v_{yz}k_y)k_z \Gamma_4 + \lambda \Gamma_5, \label{eq: continuum_e2=3}
\end{align}
where $\Gamma_i$ are the $4\times 4$ gamma matrices satisfying $\{\Gamma_i,\Gamma_j\}=2\delta_{ij}$, and $v_1,v_2,v_z,v_{xz}, v_{yz}, \lambda \in \mathbb{R}$.  
Furthermore, the $\Gamma_i$ matrices and the operator $C_{2z}\mathcal{T}$ satisfy the commutation and anticommutation relations given by  Eq.~\eqref{eq: commute_C2T_Gamm}.
The first and second terms give rise to the Euler class $|{e}_2(0)|=3$ through band inversion at $k_x=k_y=0$, since the winding number around the nodal point at $k_x=k_y=0$ becomes three due to the cubic dispersion. 
The fourth term breaks both time-reversal and $C_{2z}$ symmetries ($\mathcal{T}=K$ and $C_{2z}=1$). The fifth term is the mass term responsible for inducing a band inversion. 
We note that introducing this winding structure of both the first and the second terms renders the Euler class nontrivial [see Eq.~\eqref{eq: Euler_winding}]. Once this winding structure is imposed, the continuum Hamiltonian compatible with
$C_{2z}\mathcal{T}$ symmetry up to third order in momentum is fixed to the form of Eq.~\eqref{eq: continuum_e2=3}.

We derive an effective surface Hamiltonian from the bulk Hamiltonian in Eq.~\eqref{eq: continuum_e2=3} in a manner analogous to that in Sec.~\ref{sec: e2_2_continuum}. 
We introduce an $x$-dependence to the coefficient $\lambda$ of the mass term $\lambda \Gamma_5$ via the replacement $\lambda \rightarrow \lambda_x$, where $x$ is the distance from the surface.
The spatial profile of $\lambda_x$ is defined such that it vanishes at the surface ($\lambda_0=0$) and varies sharply to $\lambda_x=1$ for $x<0$, and to $\lambda_x=-1$ for $x>0$.  
By replacing $k_x \rightarrow -\ii \partial_x$, we obtain 
\begin{align}
    \mathcal{H}=&\ii v_1(\partial_x^3 + 3 k_y^2 \partial_x)\Gamma_1 +v_2 (-3k_y \partial_x^2 -k_y^3)\Gamma_2 \nonumber \\
    &+v_{z}k_z \Gamma_3 +(-\ii v_{xz}\partial_x + v_{yz}k_y)k_z \Gamma_4 + \lambda_x \Gamma_5. \label{eq: continuum_e2=3_2}
\end{align}

By choosing the wavevector to satisfy $3v_1k_y^2 = \pm v_{xz}k_z$, the Hamiltonian  reduces to 
\begin{align}\label{eq: Hamiltonian_ky_kz_chosen}
    \mathcal{H}_i = & - 3 v_2k_y \partial_x^2 \Gamma_2 + v_z k_z \Gamma_3  +\ii \sqrt{2}v_{xz} k_z  \partial_x \tilde{\Gamma}_i \nonumber \\
    & + \lambda_x \Gamma_5 \quad (i=1,2), 
\end{align}
where $\mathcal{H}_1$ and $\mathcal{H}_2$ correspond to $3v_1k_y^2 = v_{xz}k_z$ and $3v_1k_y^2 = -v_{xz}k_z$, respectively. Here, we retain terms up to second order in $k_y$ and neglect the third-order derivative term as a perturbation. We will discuss the case in which the third-order derivative term is dominant later in this section.
The modified gamma matrices $\tilde{\Gamma}_{i}$ ($i=1,2$) are defined as $\tilde{\Gamma}_{1}:=(\Gamma_1 - \Gamma_4)/\sqrt{2}$ and $\tilde{\Gamma}_{2}=(\Gamma_1 + \Gamma_4)/\sqrt{2}$.

The solutions $\psi_{1}$ and $\psi_{2}$ for $\mathcal{H}_1$ and $\mathcal{H}_{2}$, localized at $x=0$, are given by
\begin{align}
    &\psi_{i}=\exp \left( \frac{1}{\sqrt{2}v_{xz}k_z}\int^{x}_{0}dx'\lambda_{x'}\right) P_{i} \psi_{k_z, i},
    \label{eq: solutions_e2=3_1}
\end{align}
with $i=1,2$, where the projection operators $P_{i}$ are defined as 
\begin{gather}
P_{i}:=\frac{1}{2}\left( 1 + {\ii} \tilde{\Gamma}_i \Gamma_5 \right) \quad (i=1,2).
    \label{eq:projection_operator_e2=3}
\end{gather}
These projection operators satisfy Eq.~\eqref{eq: commutation_e2}. 
By applying the Hamiltonians $\mathcal{H}_i$ ($i=1,2$) to the eigenstates $\psi_{i}$, we obtain 
\begin{align}
    \mathcal{H}_i\psi_{i}=\exp \left( \frac{1}{\sqrt{2}v_{xz}k_z}\int^{x}_{0}dx'\lambda_{x'}\right) P_{i} \mathcal{H}_{s,i}  \psi_{k_z, i}.
\end{align}
Here, the surface Hamiltonians $\mathcal{H}_{s,i}$ ($i=1,2$) are given by
\begin{align}
        \mathcal{H}_{s,i}=P_i (  m_s \Gamma_2 +{v}_{z}k_z \Gamma_3 )P_i, \label{eq: surf_Ham_e2=3}
\end{align}
where $m_s:=-v_2 \partial_x \lambda_x / (\sqrt{2}v_1k_y)$.

Via the replacement $\partial_x \rightarrow -\partial_x$ in Eq.~\eqref{eq: continuum_e2=3_2}, we can obtain the effective surface Hamiltonian corresponding to the ($\bar{1}00$) surface. Through a similar procedure to the (100) surface, we obtain the solutions  
\begin{align}
    &\bar{\psi}_{i}=\exp \left( \frac{1}{\sqrt{2}v_{xz}k_z}\int^{x}_{0}dx'\lambda_{x'}\right) \bar{P}_{i} \bar{\psi}_{k_z, i},
    \label{eq: solutions_e2=3_2}
\end{align}
and the surface Hamiltonian 
\begin{align}
        \bar{\mathcal{H}}_{s,i}=\bar{P}_i (  -m_s \Gamma_2 +{v}_{z}k_z \Gamma_3 )\bar{P}_i, 
\end{align}
where the projection operator $\bar{P}_i$ is given by $\bar{P}_i=( 1 + {\ii} \tilde{\Gamma}_i \Gamma_5 )/2$ $(i=1,2)$.
Thus, the ($100$) and ($\bar{1}00$) surfaces host mass terms with opposite signs. 

Similarly, we can derive the $(010)$ and $(0\bar{1}0)$ surface Hamiltonians by following a procedure identical to that used for the $(100)$ and $(\bar{1}00)$ surfaces, namely, by replacing $k_y$ with $-\ii \partial_y$ in the Hamiltonian presented in Eq.~\eqref{eq: continuum_e2=3}.
By the same reasoning applied to the $(100)$ and $(\bar{1}00)$ cases, the $(010)$ and $(0\bar{1}0)$ surfaces also host mass terms with opposite signs.

In a rod geometry extending along the $z$ direction and bounded by the ($100$), ($\bar{1}00$), ($010$), and ($0\bar{1}0$) surfaces, the ($100$) and ($\bar{1}00$) surfaces host mass terms with opposite signs, as do the ($010$) and ($0\bar{1}0$) surfaces.
As discussed in Sec.~\ref{sec: e2=1}, zero-mass lines emerge between the surfaces with such surface mass terms. 
Along these zero-mass lines, the energy eigenvalues vanish at $k_z=0$.
While the argument of the exponential function in Eq.~\eqref{eq: solutions_e2=3_1} diverges at $k_z =0$, we can shift the gapless points by adding a perturbation term $\Delta \mathcal{D}$ in a manner analogous to Eq.~\eqref{eq: energyeigenvalues_with_Delta}. Thus, the two eigenstates $\psi_1$ and $\psi_2$ are gapless solutions for the surface Hamiltonian.

Since the preceding discussion did not address the case where the third-order spatial derivative dominates the spatial profiles of the boundary states, we now focus on boundary states governed by the third-order spatial derivative. By setting $k_y=0$, the Hamiltonian in Eq.~\eqref{eq: continuum_e2=3_2} reduces to 
\begin{align}
    \mathcal{H}=\ii v_1 \partial_x^3 \Gamma_1 +v_{z}k_z\Gamma_3 +\lambda_x \Gamma_5, \label{eq: e2=3_ky=0_H}
\end{align}
where we have neglected the term $-\ii v_{xz}k_z\partial_x \Gamma_4$ as a perturbation, since we focus on boundary states governed by the third-order spatial derivative.
The solutions $\psi^{(\pm)}_{3}$ for the Hamiltonian $\mathcal{H}$ in Eq.~\eqref{eq: e2=3_ky=0_H} localized at $x=0$ are given by
\begin{align}
    \psi^{(\pm)}_{3}=f_3^{(\pm)}(x) P^{(\pm)}_{3} \psi^{(\pm)}_{k_z,3},
\end{align}
where the projection operators $P^{(\pm)}_{3}$ are defined as 
\begin{gather}
P^{(\pm)}_{3}:=\frac{1}{2}\left( 1 \pm \ii {\Gamma}_1 \Gamma_5 \right).
\end{gather}
Applying the Hamiltonian $\mathcal{H}$ in Eq.~\eqref{eq: e2=3_ky=0_H} to the boundary states $\psi^{(\pm)}_{3}$ yields
\begin{align}
    \mathcal{H} \psi^{(\pm)}_{3} &= f^{(\pm)}_3(x) P_{3}^{(\pm)} v_{z}k_z \Gamma_3 \psi^{(\pm)}_{k_z,3} \nonumber \\
    &\pm \ii \Gamma_1 P^{(\pm)}_{3}(\pm v_1\partial_x^3-\lambda_x) f^{(\pm)}_3(x)\psi^{(\pm)}_{k_z,3}.
\end{align}
As discussed in Appendix~\ref{appendx: solution_DE}, exactly one of the two boundary states, either $\psi^{(+)}_{3}$ or $\psi^{(-)}_{3}$, makes the second term on the right-hand side of this equation vanish. 
Thus, the number of the boundary states governed by the third-order spatial derivative is exactly one.  

In summary, considering both Hamiltonians given in Eqs.~\eqref{eq: Hamiltonian_ky_kz_chosen} and \eqref{eq: e2=3_ky=0_H}, we obtain a total of three gapless solutions in the rod geometry. 
These three gapless modes emerge along the zero-mass line and manifest as three chiral hinge modes. 
Crucially, the first and second terms in Eq.~\eqref{eq: continuum_e2=3} drive the emergence of these hinge modes. 
Since these same terms generate the Euler class $|e_2|=3$ in the $k_x$-$k_y$ plane, this strongly suggests a direct correspondence between the three chiral hinge modes and the Euler class $|\bar{e}_2|=3$. 

\subsection{Tight-binding model}

\begin{figure}
\includegraphics[width=1.\columnwidth]{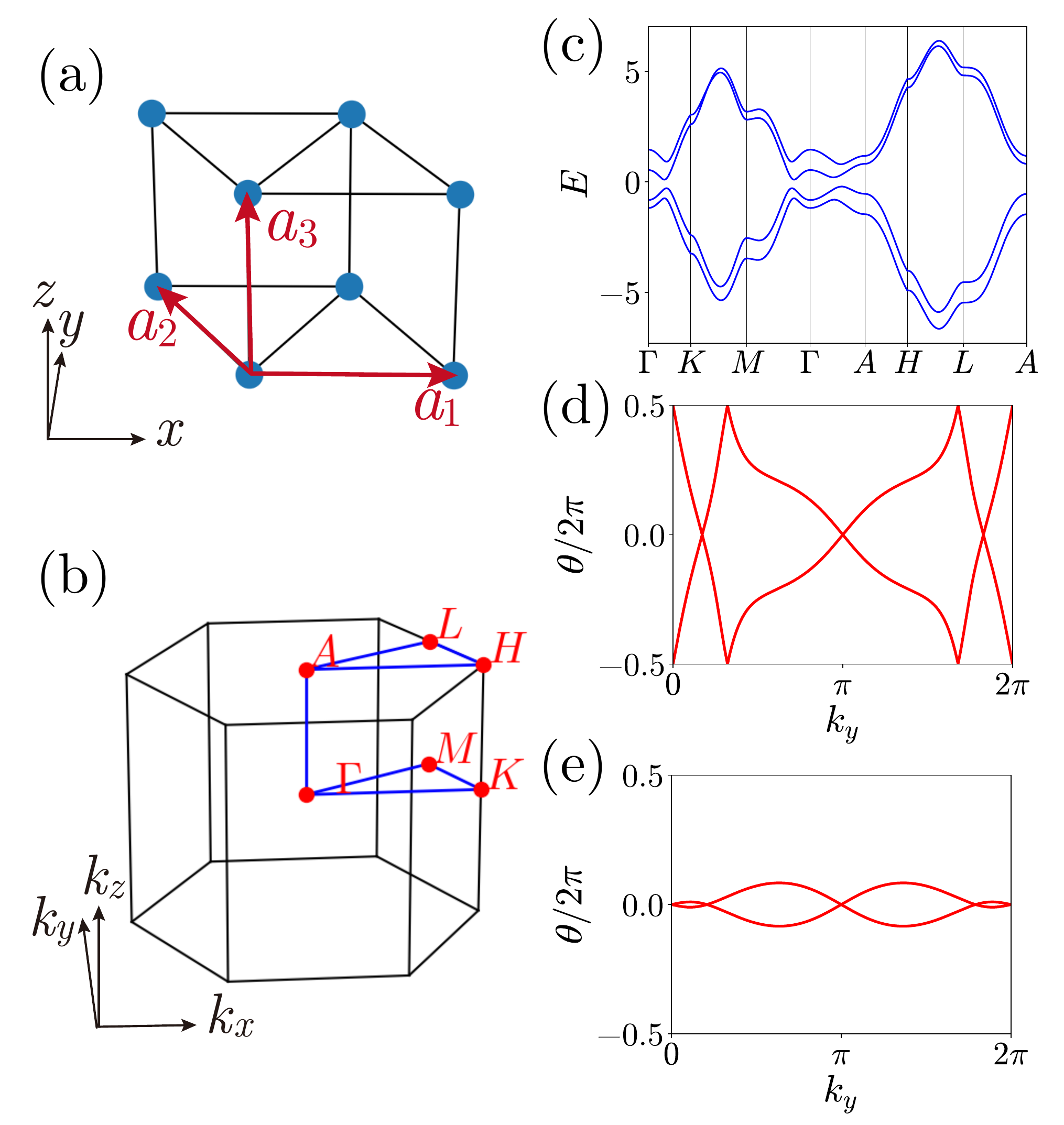}
	\caption{(a)~A three-dimensional stacked triangular lattice of the model $\mathcal{H}^{(\bar{e}_{2}=3)}_{\boldsymbol{k}}$~[Eq.~\eqref{eq:model_e2=3}]. (b)~The Brillouin zone and the high-symmetry points for the model. (c)~The bulk band structure of the model. (d,e)~The spectra of the Wilson loop operator at (d) $k_z=0$ and (e) $k_z=\pi$~[$\lambda=1$, $v_1=0.4$, $v_2=0.5$, $v_z=0.5$, $v_{xz}=4$, $v_{yz}=6$, $B_1=0.15$, $B_2=0.1$, $\Delta=0.3$]. 
 }
	\label{fig:bulk_EI_e2=3}
\end{figure}

\begin{figure*}
\includegraphics[width=2.\columnwidth]{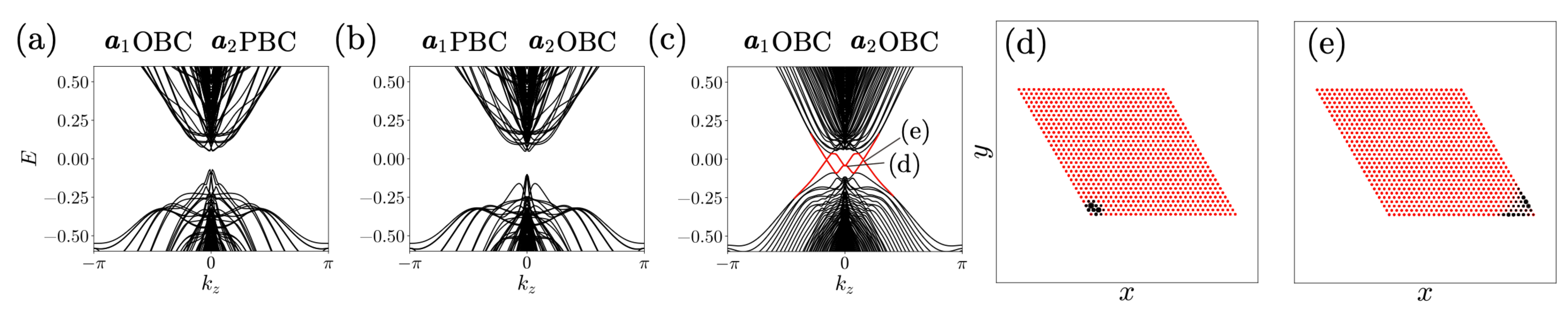}
\caption{(a-c)~Band structures of the model $\mathcal{H}^{(\bar{e}_{2}=3)}_{\boldsymbol{k}}$~[Eq.~\eqref{eq:model_e2=3}] along the $k_z$ direction under (a) the periodic boundary conditions (PBCs) both in the $\boldsymbol{a}_1$ and $z$ directions and the open boundary condition (OBC) in the $\boldsymbol{a}_2$ direction, (b) the PBCs both in the $\boldsymbol{a}_2$ and $z$ directions and the OBC in the $\boldsymbol{a}_1$ direction, (c) the OBCs both in the $\boldsymbol{a}_1$ and $\boldsymbol{a}_2$ directions and the PBC in the $z$ direction. The system sizes along the $\boldsymbol{a}_1$ and $\boldsymbol{a}_2$ directions are $L_1=30$ and $L_2=30$, respectively.
(d) and (e)~The real-space distributions of a boundary state in (c) at (d) $k_z=0.02\pi$ and $E=-0.0469$ and at (e) $k_z=0.166\pi$ and $E=-0.0150$.  Owing to the $C_{2z}\mathcal{T}$ symmetry of the system, equivalent boundary states are localized at the other hinges. Panels (d) and (e) show the distribution at one representative hinge.
The parameter values are the same as those in Fig.~\ref{fig:bulk_EI_e2=3}. 
}\label{fig:hinge_EI_e_3}
\end{figure*}

To numerically confirm our continuum theory, we employ a tight-binding model on a 3D stacked triangular lattice, where the lattice vectors are given by $\boldsymbol{a}_{1}=(1,0,0)$, $\boldsymbol{a}_2=(-1/2,\sqrt{3}/2,0)$, and $\boldsymbol{a}_3=(0,0,1)$~[Fig.~\ref{fig:bulk_EI_e2=3}(a)]. Each site hosts four orbital degrees of freedom. The four-band Bloch Hamiltonian of our model is given by 
\begin{align}
    \mathcal{H}^{(\bar{e}_{2}=3)}_{\boldsymbol{k}}=&-(4-\lambda -f_{0}(\boldsymbol{k}))\tau_z + v_1f_1(\boldsymbol{k})\sigma_x \tau_x  \nonumber \\ 
    & + v_2f_2 (\boldsymbol{k}) \sigma_z \tau_x + v_z \sin k_z \sigma_y \tau_x \nonumber \\
    &+v_{xz}\sin(\boldsymbol{k}\cdot \boldsymbol{a}_1)\sin k_z \tau_y \nonumber \\ &+\frac{v_{yz}}{\sqrt{3}}[ \sin(\boldsymbol{k}\cdot \boldsymbol{a}_2) + \sin(\boldsymbol{k}\cdot (\boldsymbol{a}_1 +\boldsymbol{a}_2)) ]\sin k_z \tau_y \nonumber  \\
    &+B_1 \sigma_x + B_2 \sigma_z + \Delta \sigma_x \tau_z, \label{eq:model_e2=3}
\end{align}
with 
\begin{align}
    f_{0} (\boldsymbol{k}) = &\cos (\boldsymbol{k}\cdot\boldsymbol{a}_{1})+\cos (\boldsymbol{k}\cdot\boldsymbol{a}_{2})+\cos (\boldsymbol{k}\cdot(\boldsymbol{a}_{1}+\boldsymbol{a}_{2})) \nonumber \\
    &+\cos (\boldsymbol{k}\cdot \boldsymbol{a}_3), \nonumber \\
    f_{1}(\boldsymbol{k})=&-8[\sin (\boldsymbol{k}\cdot\boldsymbol{a}_1)+\sin (\boldsymbol{k}\cdot \boldsymbol{a}_{2}) - \sin (\boldsymbol{k}\cdot (\boldsymbol{a}_1+\boldsymbol{a}_2))], \nonumber \\
    f_{2}(\boldsymbol{k})=&\frac{8}{3\sqrt{3}}[\sin (\boldsymbol{k}\cdot (\boldsymbol{a}_{1}+2\boldsymbol{a}_2))+ \sin (\boldsymbol{k}\cdot(-2\boldsymbol{a}_{1}-\boldsymbol{a}_2)) \nonumber \\
    &+ \sin (\boldsymbol{k}\cdot (\boldsymbol{a}_{1}-\boldsymbol{a}_2))], 
\end{align}
where $\lambda, v_1, v_2, v_z, v_{xz}, v_{yz}, B_1, B_2, \Delta \in \mathbb{R}$. 
There exist four orbitals at each lattice site, which are labeled as $\ket{\sigma_z} \otimes \ket{\tau_z}$ with $\sigma_z, \tau_z = \pm1$. This model respects $C_{2z}\mathcal{T}$ symmetry given by Eq.~\eqref{eq: C2T_symm}, where $C_{2z}\mathcal{T}$ is the complex conjugation operator. 
The high-symmetry points in the Brillouin zone for this model are shown in Fig.~\ref{fig:bulk_EI_e2=3}(b).
Figure~\ref{fig:bulk_EI_e2=3}(c) shows the bulk band structures of this model. 

To evaluate the Euler class $e_2$ in the $k_x$-$k_y$ plane at $k_z = k_z^{*}$, we introduce the $x$-directed Wilson loop matrix $\mathcal{W}(k_y,k_z^{*})$ defined by Eq.~\eqref{eq: wilson_loop}. As discussed in Sec.~\ref{sec: e2=2_lattice}, the two eigenvalues of $\mathcal{W}(k_y,k_z^{*})$ are given as a complex conjugate pair $e^{\pm \ii \theta (k_y, k_z^{*})}$.
By tracing the evolution of the phase $\theta (k_y, k_z^{*})$, we obtain the Euler class $e_2$ since the winding number of  $\theta (k_y, k_z^{*})$ is equal to the Euler class $e_2$. 
Figures~\ref{fig:bulk_EI_e2=3}(d) and \ref{fig:bulk_EI_e2=3}(e) demonstrate that the Euler class takes the values $|e_2|=3$ at $k_z=0$ and $|e_2|=0$ at $k_z=\pi$, showing the topological invariant $|\bar{e}_2|=3$. 

We numerically verify that our model $\mathcal{H}^{(\bar{e}_{2}=3)}_{\boldsymbol{k}}$ supports triple chiral hinge modes. 
Figures~\ref{fig:hinge_EI_e_3}(a) and \ref{fig:hinge_EI_e_3}(b) show the band structures under the OBC in one direction, and the PBCs in the other two directions. In these geometries, the gapless boundary states do not emerge.
Figure~\ref{fig:hinge_EI_e_3}(c) shows the band structures under the OBCs in both the $\boldsymbol{a}_1$ and $\boldsymbol{a}_2$ directions and the PBC in the $z$ direction.
In this geometry, our model supports three boundary states that are localized at the hinges~[Figs.~\ref{fig:hinge_EI_e_3}(d) and \ref{fig:hinge_EI_e_3}(e)], consistent with our continuum theory. 

\section{3D Euler insulators with $\bar{e}_2=N$}\label{sec: e2=N}
In this section, by generalizing the low-energy continuum theory for $\bar{e}_2=1$, $2$, and $3$ in Secs.~\ref{sec: e2=1}, \ref{sec: e2=2}, and \ref{sec: e2=3}, we show that a 3D Euler insulator characterized by $\bar{e}_2=N$ supports $N$ chiral hinge modes, where $N$ is a positive integer. 
We start from a generic low-energy continuum Hamiltonian 
\begin{align}
    \mathcal{H}^{(N)} = &v_1 \text{Re}(k_+^N) \Gamma_1 + v_2 \text{Im}(k_+^N) \Gamma_2 + v_z k_z \Gamma_3 \nonumber  \\
    &+ f(k_x, k_y) k_z \Gamma_4 + \lambda \Gamma_5, \label{eq: Ham_e2=N}
\end{align}
where $k_+:=k_x+\ii k_y$, and 
\begin{align}
    f(k_x, k_y) = \sum_{m \in \text{odd}}^{2 \lfloor N/2 \rfloor - 1} \left( v_{xz}^{(m)} k_x^m+v_{yz}^{(m)} k_y^m \right)
\end{align}
with $v_1, v_2, v_z, v^{(m)}_{xz}, v^{(m)}_{yz}, \lambda \in \mathbb{R}$.
The $\Gamma_i$ matrices satisfy the commutation and anticommutation relations given in Eq.~\eqref{eq: commute_C2T_Gamm}.
Here, $m$ runs over positive odd numbers, and $\lfloor x \rfloor$ denotes the floor function, which gives the greatest integer less than or equal to $x$.
The first and second terms in $\mathcal{H}^{(N)}$ give rise to the Euler class ${e}_2=N$ in the $k_x$-$k_y$ plane at $k_z=0$~\cite{Kobayashi2021PRB}. These terms are generalizations of those appearing in Eqs.~\eqref{eq: continuum_model_e2=1}, \eqref{eq: continuum_model_e2=2}, and \eqref{eq: continuum_e2=3}.
The fourth term in $\mathcal{H}^{(N)}$ is finite for $N \geq 2$ and breaks both time-reversal and $C_{2z}$ symmetries, where $\mathcal{T}=K$ and $C_{2z}=1$. 
Using the binomial theorem, the real and imaginary parts of $k_+^N$ can be expressed as 
\begin{gather}
\text{Re}(k_+^N) = \sum_{j=0}^{\lfloor N/2 \rfloor} \binom{N}{2j} (-1)^j k_x^{N-2j} k_y^{2j}, \label{eq: Rek_+N} \\
\text{Im}(k_+^N) = \sum_{j=0}^{\lfloor (N-1)/2 \rfloor} \binom{N}{2j+1} (-1)^j k_x^{N-2j-1} k_y^{2j+1}. \label{eq: Imk_+N}
\end{gather}

Before discussing the even and odd $N$ cases separately, we explain why they require distinct treatments. Upon the substitution $k_x\to-\ii \partial_x$, the parities of the spatial derivatives generated by the $v_1\mathrm{Re}(k_+^N)\Gamma_1$ and $v_2\mathrm{Im}(k_+^N)\Gamma_2$ terms depend on the parity of $N$, as seen from Eqs.~\eqref{eq: Rek_+N} and \eqref{eq: Imk_+N}. Because the gapless boundary states arise from the odd-order spatial derivatives (see Sec.~\ref{sec: e2_2_continuum} and Sec.~\ref{sec: e2=3_continuum}), the roles played by the $v_1\mathrm{Re}(k_+^N)\Gamma_1$ and $v_2\mathrm{Im}(k_+^N)\Gamma_2$ terms are interchanged between even and odd $N$. For these reasons, we treat the even and odd $N$ cases separately in the following.

\subsection{even $N$}\label{sec: e2=N_evenN}
When $N$ is even, in a manner analogous to Sec.~\ref{sec: e2_2_continuum}, we show that the Hamiltonian given in Eq.~\eqref{eq: Ham_e2=N} hosts $N$ gapless boundary states on the (100) surface by substituting $k_x \rightarrow -\ii \partial_x$. 
We focus on the case where the $m$-th order spatial derivative gives rise to gapless boundary states, where $m$ is an odd number. 
The following discussion applies to any odd $m$, which takes values $m=1,3, \dots, N-1$. 
Which $m$-th order spatial derivative plays the dominant role in generating the gapless boundary states depends on the values of $k_y$ and $\partial_x \sim 1/\xi$, where $\xi$ is the penetration length of the boundary states.
When the wavevector is chosen to satisfy 
\begin{align}
v_{2} \binom{N}{N-m} (-1)^{\frac{N-m-1}{2}}k_y^{N-m}=\pm v^{(m)}_{xz}k_z, \label{eq: e_2=N_i=1_cond}  
\end{align}
 the Hamiltonian~[Eq.~\eqref{eq: Ham_e2=N}] reduces to 
\begin{align}
    \mathcal{H}^{(N)}_{i}& = v_1 \sum_{j=0}^{ N/2} \binom{N}{2j} (-1)^j k_y^{2j} (-\ii \partial_x)^{N-2j} \Gamma_1  \nonumber \\
    &+v_2\sum_{l \in \text{odd}, l \neq m}^{N-1} \binom{N}{N-l} (-1)^{\frac{N-l-1}{2}} k_y^{N-l} (-\ii \partial_x)^{l} \Gamma_2 \nonumber \\
    &+ \sum_{l \in \text{odd},l\neq m}^{N- 1}  v_{xz}^{(l)} k_z (-\ii \partial_x)^l \Gamma_4  + \sum_{l \in \text{odd}}^{N - 1}v_{yz}^{(l)} k_y^l k_z \Gamma_4 \nonumber \\
    &+v_z k_z \Gamma_3+\sqrt{2}v_{xz}^{(m)} k_z (-\ii\partial_x)^m \tilde{\Gamma}_i+\lambda \Gamma_5, \label{eq: Ham_e2=N_2}
\end{align}
for $i=1,2$, where $l$ runs over positive odd numbers, and $\mathcal{H}^{(N)}_{1}$ and $\mathcal{H}^{(N)}_{2}$ correspond to the $+$ and $-$ signs in Eq.~\eqref{eq: e_2=N_i=1_cond}, respectively. 
Here, $\tilde{\Gamma}_i$ ($i=1,2$) are defined as $\tilde{\Gamma}_1=(\Gamma_2+ \Gamma_4)/\sqrt{2}$ and $\tilde{\Gamma}_2=(-\Gamma_2+ \Gamma_4)/\sqrt{2}$. 

To capture the gapless boundary modes, we focus on a regime characterized by the penetration length $\xi \sim 1/|\partial_x|$ of the boundary states, where the $m$-th order spatial derivative dominates their spatial profiles. 
Under this condition, spatial derivative terms of order $l\neq m$ become parametrically small and can be treated as perturbations. Furthermore, since the momentum-only term $\sum_{l \in \text{odd}}^{N - 1}v_{yz}^{(l)} k_y^l k_z \Gamma_4$ represents higher-order corrections in momentum relative to the leading-order linear term $v_z k_z \Gamma_3$, we neglect it perturbatively. We also neglect the first term in Eq.~\eqref{eq: Ham_e2=N_2}, since it consists of the even order spatial derivatives that induce surface mass terms, as discussed in Sec.~\ref{sec: e2_2_continuum} [see Eq.~\eqref{eq:surface_dirac_Hamiltonian2} and Eq.~\eqref{eq: surface_100}]. 
Consequently, the Hamiltonians $\mathcal{H}^{(N)}_{i,m}$ ($i=1,2$) governed by the $m$-th order spatial derivative are given by
\begin{align}
    \mathcal{H}^{(N)}_{i,m} = v_z k_z \Gamma_3+\sqrt{2}v_{xz}^{(m)} k_z (-\ii \partial_x)^m \tilde{\Gamma}_i+\lambda \Gamma_5. \label{eq: mth_order_H}
\end{align}
The solutions $\psi^{(\pm)}_{1,m}$ and $\psi^{(\pm)}_{2,m}$ for the Hamiltonians $\mathcal{H}^{(N)}_{1,m}$  and $\mathcal{H}^{(N)}_{2,m}$, respectively, localized at $x=0$ are given by
\begin{align}\label{eq: e2=N_eigenstates}
    \psi^{(\pm)}_{i,m}=f^{(\pm)}_{i,m}(x) P^{(\pm)}_{i,m} \psi^{(\pm)}_{k_z, i,m},
\end{align}
with $i=1,2$, where the projection operators $P^{(\pm)}_{i,m}$ are defined as 
\begin{gather}
P^{(\pm)}_{i,m}:=\frac{1}{2}\left( 1 \mp \ii^m \tilde{\Gamma}_i \Gamma_5 \right) \quad (i=1,2).
\end{gather}

Applying the Hamiltonian $\mathcal{H}^{(N)}_{i,m}$ to the eigenstate $\psi^{(\pm)}_{i,m}$ yields
\begin{align}
    &\mathcal{H}^{(N)}_{i,m}\psi^{(\pm)}_{i,m}=f^{(\pm)}_{i,m}(x)P^{(\pm)}_{i,m}v_z k_z \Gamma_3 \psi^{(\pm)}_{k_z, i,m} \nonumber  \\
    &\quad  \mp\ii^{m}\tilde{\Gamma}_i P^{(\pm)}_{i,m} \left( \pm \sqrt{2}v_{xz}^{(m)}k_z \partial_x^m  -\lambda  \right) f^{(\pm)}_{i,m}(x)\psi^{(\pm)}_{k_z, i,m}. \label{eq: mth_order_DE}
\end{align}
As discussed in Appendix~\ref{appendx: solution_DE}, provided that the parameter $\lambda$ is given by $\lambda = - \text{sgn}(x)$, exactly one of the two boundary states, either $\psi^{(+)}_{i,m}$ or $\psi^{(-)}_{i,m}$, makes the second term on the right-hand side of Eq.~\eqref{eq: mth_order_DE} vanish. This yields one boundary state for each of the Hamiltonians $\mathcal{H}^{(N)}_{1,m}$ and $\mathcal{H}^{(N)}_{2,m}$, resulting in a total of two boundary states for a given $m$. 
Since $m$ takes values $m=1,3,\dots, N-1$, the total number of gapless boundary states is $N$. 
These $N$ boundary states give rise to $N$ chiral hinge modes because the mass term originating from the first term in Eq.~\eqref{eq: Ham_e2=N_2} takes opposite signs on the (100) and $(\bar{1}00)$ surfaces, as discussed in Sec.~\ref{sec: e2_2_continuum}.

\subsection{odd $N$}
When $N$ is odd, in a manner analogous to Sec.~\ref{sec: e2=3_continuum}, we show that the Hamiltonian given in Eq.~\eqref{eq: Ham_e2=N} hosts $N$ gapless boundary states on the (100) surface by substituting $k_x \rightarrow -\ii \partial_x$.
Similar to the even-$N$ case, we focus on the case where the $m$-th order spatial derivative gives rise to gapless boundary states, where $m$ is an odd number. The following discussion applies to any odd $m$, which takes values $m=1,3, \dots, N-2$. 
When the wavevector is chosen to satisfy
\begin{align}
v_{1} \binom{N}{N-m} (-1)^{\frac{N-m}{2}}k_y^{N-m}=\pm v^{(m)}_{xz}k_z, \label{eq: e_2=N_cond2}  
\end{align}
 the Hamiltonian~[Eq.~\eqref{eq: Ham_e2=N}] reduces to 
\begin{align}
    \mathcal{H}^{(N)}_{i} &= v_1 \sum_{l \in \text{odd}, l \neq m}^{N} \binom{N}{N-l} (-1)^{\frac{N-l}{2}} k_y^{N-l} (-\ii \partial_x)^{l} \Gamma_1  \nonumber \\
    &+v_2\sum_{l \in \text{odd}}^{N} \binom{N}{l} (-1)^{\frac{l-1}{2}} k_y^{l} (-\ii \partial_x)^{N-l} \Gamma_2 \nonumber \\
    &+ \sum_{l \in \text{odd},l\neq m}^{N-2}  v_{xz}^{(l)} k_z (-\ii \partial_x)^l\Gamma_4  + \sum_{l \in \text{odd}}^{N-2}v_{yz}^{(l)} k_y^l k_z \Gamma_4 \nonumber \\
    &+v_z k_z \Gamma_3+\sqrt{2}v_{xz}^{(m)} k_z (-\ii\partial_x)^m \tilde{\Gamma}_i+\lambda \Gamma_5, \label{eq: Ham_e2=N_odd}
\end{align}
for $i=1,2$, where $l$ runs over positive odd numbers, and $\mathcal{H}^{(N)}_1$ and $\mathcal{H}^{(N)}_2$ correspond to the $+$ and $-$ signs in Eq.~\eqref{eq: e_2=N_cond2}, respectively.
Here, $\tilde{\Gamma}_i$ ($i=1,2$) are defined as $\tilde{\Gamma}_1=(\Gamma_1+ \Gamma_4)/\sqrt{2}$ and $\tilde{\Gamma}_2=(-\Gamma_1+ \Gamma_4)/\sqrt{2}$. 

Similar to the even-$N$ case, we focus on the case where the $m$-th order spatial derivative dominates the spatial profiles of the boundary states, and we drop the first four terms in Eq.~\eqref{eq: Ham_e2=N_odd}. Consequently, the Hamiltonians $\mathcal{H}^{(N)}_{i,m}$ ($i=1,2$) governed by the $m$-th order spatial derivative are given by Eq.~\eqref{eq: mth_order_H} in Sec.~\ref{sec: e2=N_evenN}. 
Applying the Hamiltonian $\mathcal{H}_{i,m}^{(N)}$ to the eigenstates $\psi^{(\pm)}_{i,m}$ given by Eq.~\eqref{eq: e2=N_eigenstates} yields Eq.~\eqref{eq: mth_order_DE}.
The number of boundary states that make the second term on the right-hand side of Eq.~\eqref{eq: mth_order_DE} vanish is exactly one (see Appendix~\ref{appendx: solution_DE}) for each of the Hamiltonians $\mathcal{H}_{1,m}^{(N)}$ and $\mathcal{H}_{2,m}^{(N)}$, resulting in a total of two boundary states for a given $m$.
Since $m$ takes values $m=1,3,\dots, N-2$, the total number of gapless boundary states is $N-1$.

Since the preceding discussion did not include the case where the $N$-th order spatial derivative dominates the spatial profiles of the boundary states, we now examine the Hamiltonian $\mathcal{H}^{(N)}_{N}$ governed by this $N$-th order spatial derivative.
By setting $k_y=0$, the Hamiltonian in Eq.~\eqref{eq: Ham_e2=N_odd} reduces to
\begin{align}
    \mathcal{H}^{(N)}_{N} = v_1(-\ii \partial_x)^{N}\Gamma_1+ v_z k_z \Gamma_3+ \lambda \Gamma_5,
\end{align}
where we have neglected the term $\sum_{l \in \text{odd}}^{N-2}  v_{xz}^{(l)} k_z (-\ii \partial_x)^l\Gamma_4$ as a perturbation.
The solutions $\psi^{(\pm)}_{N}$ for the Hamiltonian $\mathcal{H}^{(N)}_{N}$ localized at $x=0$ are given by
\begin{align}
    \psi^{(\pm)}_{N}=f^{(\pm)}_{N}(x) P^{(\pm)}_{N} \psi^{(\pm)}_{k_z,N},
\end{align}
where the projection operators $P^{(\pm)}_{N}$ are defined as 
\begin{gather}
P^{(\pm)}_{N}:=\frac{1}{2}\left( 1 \mp \ii^N {\Gamma}_1 \Gamma_5 \right).
\end{gather}
Applying the Hamiltonian $\mathcal{H}^{(N)}_{N}$ to the boundary states $\psi^{(\pm)}_{N}$ yields
\begin{align}
    \mathcal{H}^{(N)}_{N} \psi^{(\pm)}_{N} &= f^{(\pm)}_N(x) P_{N}^{(\pm)} v_{z}k_z \Gamma_3 \psi^{(\pm)}_{k_z,N} \nonumber \\
    &\mp\ii^N \Gamma_1 P^{(\pm)}_{N}(\pm v_1\partial_x^N-\lambda) f^{(\pm)}_N(x)\psi^{(\pm)}_{k_z,N}.
\end{align}
As shown in Appendix~\ref{appendx: solution_DE}, one of the two boundary states, either $\psi^{(+)}_{N}$ or $\psi^{(-)}_{N}$, makes the second term on the right-hand side vanish.
In summary, by considering the Hamiltonian $\mathcal{H}^{(N)}_{N}$ governed by the $N$-th order spatial derivative in addition to the Hamiltonians $\mathcal{H}^{(N)}_{m}$ in Eq.~\eqref{eq: mth_order_H} with $m=1,3,\dots, N-2$, we obtain a total of $N$ gapless boundary states.  

\section{Conclusions and discussion}\label{sec: conclusions_discussion}
In this work, we have studied 3D Euler insulators characterized by the topological invariant $\bar{e}_2=e_2(0)-e_2(\pi)$, defined as the difference in the Euler class $e_2$ between the $k_z=0$ and $k_z=\pi$ planes, and demonstrated the emergence of multiple chiral hinge modes characterized by this invariant. Focusing on systems with $C_{2z}\mathcal{T}$ symmetry, we constructed generic low-energy continuum Hamiltonians for $\bar{e}_2 = 1$, $2$, and $3$ and derived the corresponding surface Hamiltonians. We showed that sign changes in the surface mass lead to domain walls supporting multiple chiral hinge modes. 
We numerically verified these predictions using 3D tight-binding models on simple cubic and stacked triangular lattices, confirming the presence of multiple hinge modes consistent with the continuum theory. Finally, we generalized the continuum theory to arbitrary $\bar{e}_2 = N$ and demonstrated that a 3D Euler insulator characterized by $\bar{e}_2 = N$ supports $N$ chiral hinge modes.

Given the experimental realizations of Euler band topology in various artificial platforms, such as acoustic metamaterials~\cite{Peri2020Sience, Jiang2021NatPhys, Qiu2023, Jian2024}, photonic crystals~\cite{Slobozhanyuk2017three, Park2021ACS, Kim2022three, Wang2022PRL, Hu2024NatCommun, Yi2024Science, Wenwen2025SciAdv, wang2025realization}, and transmission
line networks~\cite{Guo2021Nature, Jiang2021NatureCommun},  exploring the experimental realization of 3D Euler insulators supporting these multiple hinge modes in these platforms is a promising direction.
Furthermore, Euler band topology often also emerges in electronic systems, such as twisted bilayer graphene~\cite{Ahn2019PRX}, ZrTe~\cite{Bouhon2020}, ZrTe$_5$~\cite{Lee2025PRB}, RE$_8$CoX$_3$ family (RE $=$ rare earth elements, X $=$ Al, Ga, or In)~\cite{Sato2024}, and  superfluid $^3$He-B~\cite{Kobayashi2026PRB}. 
Although isolating a pair of real bands from other bands in 3D insulators is generally challenging, multiple hinge modes are nevertheless expected to emerge in realistic electronic systems. In particular, when the Euler class is odd, at least one chiral hinge mode remains robust regardless of the number of bands, due to its connection to the Chern-Simons invariant.

\begin{acknowledgments}
This work was supported by the Japan Society for the Promotion of Science (JSPS) KAKENHI Grants No.~JP24K22868, No.~JP24K00557, and No.~JP25K07161 and by JST CREST Grant No.~JPMJCR19T2.
\end{acknowledgments}

\appendix

\section{Derivation of the surface Hamiltonian from the lattice model}\label{appendix: derive_surf_Ham}
In this appendix, we derive an effective surface Hamiltonian from the lattice model in Eq.~\eqref{eq:lattice_model}. 
Following a procedure similar to the derivation of the surface Hamiltonian in Sec.~\ref{sec: e2_2_continuum}, we
expand this Hamiltonian around $\boldsymbol{k}=0$. 
Substituting $k_x \rightarrow -\ii \partial_x$ and $-(3-\lambda -\sum_{i=x,y,z}\cos k_i) \rightarrow \lambda_{x}$, we obtain 
\begin{align}
     \mathcal{H}^{(\bar{e}_{2}=2)}_{\boldsymbol{k}} =&v_1 \sigma_x \tau_x \partial_x^2 -\ii (v_2 k_y \sigma_z \tau_x +v_{xz}k_z \tau_y \sigma_x) \partial_x \nonumber \\
    &+v_z k_z \sigma_y \tau_x + \lambda_x \tau_z, 
\end{align}
where we neglect terms quadratic in the wavevector as well as the perturbation terms $B_1\sigma_x$, $B_2 \sigma_z$, and $\Delta \tau_x$.
Here, the spatial profile of $\lambda_x$ is defined such that it vanishes at the surface ($\lambda_0=0$) and varies sharply to $\lambda_x=1$ for $x<0$, and to $\lambda_x=-1$ for $x>0$. 
By setting $v_2 k_y= v_{xz}k_z$ or $v_2 k_y= -v_{xz}k_z$, we obtain the Hamiltonians
\begin{align}
     \mathcal{H}^{(\bar{e}_{2}=2)}_{\boldsymbol{k},1} =&v_1 \sigma_x \tau_x \partial_x^2 -\ii v_{xz}k_z(\sigma_z \tau_x + \tau_y) \partial_x \nonumber \\
    &+v_z k_z \sigma_y \tau_x + \lambda_x \tau_z , \nonumber \\
     \mathcal{H}^{(\bar{e}_{2}=2)}_{\boldsymbol{k},2} =&v_1 \sigma_x \tau_x \partial_x^2 -\ii v_{xz}k_z(-\sigma_z \tau_x +  \tau_y) \partial_x \nonumber \\
    &+v_z k_z \sigma_y \tau_x + \lambda_x \tau_z,    
\end{align}
where $\mathcal{H}^{(\bar{e}_{2}=2)}_{\boldsymbol{k},1}$ and $\mathcal{H}^{(\bar{e}_{2}=2)}_{\boldsymbol{k},2}$ correspond to $v_2 k_y= v_{xz}k_z$ and $v_2 k_y= -v_{xz}k_z$, respectively.
We obtain the solutions $\psi^{(\pm)}_{1}$ and $\psi^{(\pm)}_{2}$, as defined by
\begin{align}
    &\psi^{(\pm)}_{i}=\exp \left( \frac{\pm 1}{\sqrt{2}v_{xz}k_z}\int^{x}_{0}dx'\lambda_{x'}\right) P^{(\pm)}_{i} \psi^{(\pm)}_{k_z, i} \ \ (i=1,2), 
\end{align}
for the Hamiltonians $\mathcal{H}^{(\bar{e}_{2}=2)}_{\boldsymbol{k},1}$  and $\mathcal{H}^{(\bar{e}_{2}=2)}_{\boldsymbol{k},2}$, respectively. The projection operators $P_{i}^{(\pm)}$ ($i=1,2$) are defined as 
\begin{gather}
P_{1}^{(\pm)}:=\frac{1}{2}\left( 1 \mp \frac{1}{\sqrt{2}} (\sigma_z \tau_y - \tau_x) \right), \nonumber \\
P_{2}^{(\pm)}:=\frac{1}{2}\left( 1 \mp \frac{1}{\sqrt{2}}(-\sigma_z \tau_y - \tau_x) \right).
\end{gather}

Applying the Hamiltonian $\mathcal{H}^{(\bar{e}_{2}=2)}_{\boldsymbol{k}, i}$ ($i=1,2$) to the states $\psi^{(\pm)}_{i}$ in a manner analogous to Eq.~\eqref{eq: H_psi}, we obtain the effective surface Hamiltonian
\begin{align}
    \mathcal{H}^{(\pm)}_{s,i}=P^{(\pm)}_{i}[ v_z k_{z}\sigma_y \tau_x + m^{(\pm)} \sigma_x \tau_x  ]P^{(\pm)}_{i},\label{eq: app_surface_100}
\end{align}
with
\begin{align}
    m^{(\pm)}:= \pm \frac{v_1 \partial_x \lambda_x}{\sqrt{2}v_{xz}k_z},
\end{align}
where we have neglected terms of order $\mathcal{O}(\lambda_x^2)$, assuming $\lambda_x \ll 1$ near the surface.
To make the formulation more transparent, it is convenient to express the projection operators $P^{(\pm)}_{i}$ in the form $(1 \pm \tau_z)/2$. Thus, we perform the following unitary transformation:
\begin{gather}\label{eq:unitary_transform}
    U_1:=\frac{1}{\sqrt{2}}\left[ 1+\frac{\ii}{\sqrt{2}}(\sigma_z \tau_x + \tau_y)\right], \\
    U_2:=\frac{1}{\sqrt{2}}\left[ 1-\frac{\ii}{\sqrt{2}}(\sigma_z \tau_x - \tau_y)\right],    
    \\ P'^{(\pm)}_{i}=U_i^{\dagger}P^{(\pm)}_i U_i = \frac{1}{2}(1 \mp \tau_z), \quad (i=1,2).
\end{gather}
Under this unitary transformation, the surface Hamiltonians $\mathcal{H}^{(\pm)}_{s,i}$ in Eq.~\eqref{eq: app_surface_100} become
\begin{align}
\mathcal{H}'^{(\pm)}_{s,1}:=&P'^{(\pm)}_{1}U_1^{\dagger}\mathcal{H}^{(\pm)}_{s,1}U_1P'^{(\pm)}_{1} \nonumber \\
    =&v_zk_z\frac{-\sigma_x \pm \sigma_y}{\sqrt{2}}+ m^{(\pm)} \frac{\pm \sigma_x +\sigma_y}{\sqrt{2}},  \\
\mathcal{H}'^{(\pm)}_{s,2}:=&P'^{(\pm)}_{2}U_2^{\dagger}\mathcal{H}^{(\pm)}_{s,2}U_2P'^{(\pm)}_{2} \nonumber \\
    =&v_zk_z\frac{\sigma_x \pm \sigma_y}{\sqrt{2}}+ m^{(\pm)} \frac{\pm \sigma_x -\sigma_y}{\sqrt{2}}.    
\end{align}
Here, the degrees of freedom $\tau_z =\pm1$ correspond to the top and bottom surfaces, and therefore $\mathcal{H}'^{(+)}_{s,i}$ is the surface Hamiltonian for one surface, while $\mathcal{H}'^{(-)}_{s,i}$ corresponds to the opposite surface. 
The energy eigenvalues for both of these Hamiltonians are given by
\begin{align}
    E= \pm \sqrt{(v_z k_z)^2+(m^{(\pm)})^2},
\end{align}
and the $(100)$ and $(\bar{1}00)$ surfaces host mass terms $m^{(+)}$ and $m^{(-)}$ with opposite signs. 
As discussed in the main text, such mass terms with opposite signs give rise to chiral hinge modes.
Consequently, two chiral hinge modes associated with the $\mathcal{H}'^{(+)}_{s,1}$ and $\mathcal{H}'^{(+)}_{s,2}$ appear at the line where the mass vanishes.

\section{The number of gapless boundary states governed by $m$-th order spatial derivative}\label{appendx: solution_DE}

In this appendix, we show that the number of boundary states localized around $x=0$ with a spatial profile $f(x)$ is exactly one for only one of the two cases: $A>0$ or $A<0$.
Here, $f(x)$ is a function satisfying 
\begin{align}
    (A\partial_x^m - \lambda)f(x)=0, \label{eq: ap_m_order_DE}
\end{align}
where $A$ is a nonzero real parameter, $m$ is a positive odd integer, and $\lambda$ is given by $\lambda=-\text{sgn}(x)$.
We assume a solution of the form $f(x)=e^{rx}$ for Eq.~\eqref{eq: ap_m_order_DE}, where $r$ is a complex number.
Introducing a complex variable $z$ satisfying $z^m=Ar^m$, the characteristic equation for Eq.~\eqref{eq: ap_m_order_DE} is written as
\begin{gather}
    z^m = \begin{cases}
        -1, \quad  (x>0), \\
        1, \quad  (x<0).
    \end{cases}
\end{gather}
For $z^m=-1$, let $p$ and $q$ denote the number of solutions satisfying $\text{Re}(z)>0$ and $\text{Re}(z)<0$, respectively.
For $z^m=1$, it follows that the number of solutions satisfying $\text{Re}(z)>0$ and $\text{Re}(z)<0$ are $q$ and $p$, respectively, because $m$ is an odd integer.
Furthermore, since $p$ and $q$ satisfy $p+q=m$ and $m$ is odd, the pair of $(p,q)$ is given by
\begin{align}
    (p,q)=\left( \frac{m+1}{2}, \frac{m-1}{2} \right), \label{eq: p_q_value1}
\end{align} 
or
\begin{align}
    (p,q)=\left( \frac{m-1}{2}, \frac{m+1}{2} \right). \label{eq: p_q_value2}
\end{align}

\subsection{$A>0$}
In this subsection, we discuss the case where $A>0$.
\begin{itemize}
    \item For $x>0$ ($z^m=-1$), we consider solutions that satisfy the boundary condition $f(x)\rightarrow 0$ as $x \rightarrow \infty$. In this case, $r$ must satisfy $\text{Re}[r]<0$, which implies $\text{Re}[z]<0$. The number of solutions satisfying both $z^m=-1$ and $\text{Re}[z]<0$ is $q$.
    \item Similarly, for $x<0$ ($z^m=1$), we require the solutions to satisfy $f(x)\rightarrow 0$ as $x \rightarrow -\infty$. In this case, $r$ must satisfy $\text{Re}[r]>0$, which implies $\text{Re}[z]>0$. The number of solutions satisfying both $z^m=1$ and $\text{Re}[z]>0$ is also $q$.
\end{itemize}
Consequently, we construct the general solution $f(x)$ by taking a linear combination of these $2q$ solutions of the form $e^{rx}$. Because of the discontinuity of $\lambda = -\text{sgn}(x)$ at $x=0$, the function $f(x)$ and its derivatives up to the $(m-1)$-th order ($f, \partial_x f, \partial_x^2 f, \dots, \partial_x^{m-1}f$) must be continuous at $x=0$. Imposing these $m$ constraint conditions, the number of independent solutions becomes $2q-m.$

\subsection{$A<0$}

In this subsection, we discuss the case where $A<0$.
\begin{itemize}
    \item For $x>0$ ($z^m=-1$), we consider solutions that satisfy the boundary condition $f(x)\rightarrow 0$ as $x \rightarrow \infty$. In this case, $r$ must satisfy $\text{Re}[r]<0$, which implies $\text{Re}[z]>0$. The number of solutions satisfying both $z^m=-1$ and $\text{Re}[z]>0$ is $p$.
    \item Similarly, for $x<0$ ($z^m=1$), we require the solutions to satisfy $f(x)\rightarrow 0$ as $x \rightarrow -\infty$. In this case, $r$ must satisfy $\text{Re}[r]>0$, which implies $\text{Re}[z]<0$. The number of solutions satisfying both $z^m=1$ and $\text{Re}[z]<0$ is also $p$.
\end{itemize}
Consequently, we construct the general solution $f(x)$ by taking a linear combination of these $2p$ solutions of the form $e^{rx}$. The function $f(x)$ and its derivatives up to the $(m-1)$-th order must be continuous at $x=0$. Imposing these $m$ constraint conditions, the number of independent solutions becomes $2p-m.$

\subsection{The number of solutions}
If $p$ and $q$ satisfy Eq.~\eqref{eq: p_q_value1}, exactly one solution exists for $A<0$ since $2p-m=1$, whereas no solution exists for $A>0$ since $2q-m=-1<0$.
Conversely, if $p$ and $q$ satisfy Eq.~\eqref{eq: p_q_value2}, exactly one solution exists for $A>0$ since $2q-m=1$, whereas no solution exists for $A<0$ since $2p-m=-1<0$.
Therefore, regardless of whether $(p,q)$ satisfies Eq.~\eqref{eq: p_q_value1} or Eq.~\eqref{eq: p_q_value2},
we obtain exactly one solution for only one of the two cases: $A>0$ or $A<0$.

%\bibliography{reference}

%

%\begin{comment}

%\end{comment}

\end{document}